\documentclass[runningheads,citeauthoryear]{apinv}
\usepackage{epsfig,cite,graphics}
\usepackage{marvosym} 
\usepackage[utf8]{inputenc}

\begin{document}

\title{ESpeRo: Echelle Spectrograph Rozhen}
\titlerunning{ESpeRo: Echelle Spectrograph Rozhen}
\author{
Tanyu Bonev\inst{1},
Haralambi Markov\inst{1},
Toma Tomov\inst{2},\\
Rumen Bogdanovski\inst{1},
Pencho Markishki\inst{1},
Maya Belcheva\inst{1},\\
Wojciech Dimitrov\inst{3},
Krzysztof Kami{\'n}ski\inst{3},\\
Ilko Milushev\inst{4},
Faig Musaev\inst{5}, 
Mirela Napetova\inst{1},\\
Grigor Nikolov\inst{1}, 
Plamen Nikolov\inst{1}, 
Tihomir Tenev\inst{4} 
}
\authorrunning{T. Bonev, H. Markov, T. Tomov, et al.}
\tocauthor{T. Bonev, H. Markov, T. Tomov, et al.} 
\institute{
Institute of Astronomy and NAO, Bulgarian Academy of Sciences, BG-1784, Sofia 
\and Centre for Astronomy, Faculty of Physics, Astronomy and Informatics, Nicolaus Copernicus University, Grudziadzka 5, PL-87-100 Torun, Poland
\and Astronomical Observatory Institute, Faculty of Physics, A. Mickiewicz University,
S{\l}oneczna   36, 60-286 Pozna\'{n}, Poland
\and  Georgi Nadjakov Institute of Solid State Physics,  Bulgarian Academy of Sciences, BG-1784, Sofia
\and Special Astrophysical Observatory of the Russian AS, Nizhnij Arkhyz, 369167, Russia; Institute of Astronomy of the Russian AS, 48 Pyatnitskaya st., 119017, Moscow, Russia; Terskol Branch of Institute of Astronomy of the Russian AS, 361605 Peak Terskol, Kabardino-Balkaria, Russia
\newline
\email{tbonev@astro.bas.bg} 
   }
\papertype{Submitted on xx.xx.xxxx; Accepted on xx.xx.xxxx}	
\maketitle

\begin{abstract}

In this paper we describe the echelle spectrograph of the 2 meter telescope of the Rozhen National Astronomical Observatory. The spectrograph is a cross-dispersed, bench-mounted, fiber-fed instrument giving a resolution from $\sim$30000 to $\sim$45000. The spectral range obtained in one single image is from 3900 to 9000 {\AA}. We describe the parameters of the fiber injection and the guiding unit, of the spectrograph itself, and of the detector. The identified orders and the resulting resolving power are presented. The opportunity to increase of the resolution by using a narrower slit is discussed and the corresponding loss of flux is calculated. The expected signal-to-noise ratio for a set of stars of different magnitudes was derived. Some of the first results obtained with ESpeRo are shortly described.
\end{abstract}

\section*{Introduction}

For more than 35 years the spectroscopic research of Bulgarian astronomers has been mainly based on data obtained with the high-quality Coude spectrograph, specifically designed for the 2 meter telescope of the National Astronomical Observatory Rozhen (NAO). Description of the basic characteristics of the Coude spectrograph can be found in Kolev (2008) and at the NAO webpage\footnote{http://www.nao-rozhen.org}. The spectrograph was designed in the 1970s when the common standard detectors were the photoplates. Based on the data obtained with the Coude spectrograph, a great number of original results have been published in different fields of research, during the era of photoplates. It is not possible to mention all of them, but it is worthed to cite at least several papers, mainly in order to illustrate the process of transition from the era of photoplates to CCDs. Properties of the stellar wind have been analysed by Markova (1986), Tomov et al. (1990) explained the strange behaviour of the unique star MWC 560, peculiar profiles of hydrogen lines were found in the spectra of several $\lambda$ Bootis stars, investigated by Iliev and Barzova (1993), Stateva (1995) explained the helium distribution on the surface of the CP4 star HD 21699. Immediately after changing from photoplates to CCDs, it was clear that we need an echelle spectrograph to continue doing good science with the 2-m telescope. 
Markov (2011) made a summary of the scientific motivation and a review of possible solutions for an echelle spectrograph for the 2-meter telescope of NAO.

The CCDs were introduced in the Coude spectrograph in 1993 and today an 1Kx1K Photometrics camera is used. The CCD highly improved the spectrograph light efficiency and the spectral quality (signal-to-noise ratio) at the cost of the reduced spectral range available with a single exposure. Originally designed to cover a spectral window of ~1000 and ~1500 {\AA} on a photographic plate (using respective gratings) the CCD image nowadays covers spectral ranges of 200 and 100 {\AA} with a resolving power (R) of 15000 and 30000 respectively. Evidently, with a medium-sized CCD detector the Coude spectrograph offers about an order of magnitude smaller spectral range. But most of the science cases require acquisition of the entire spectral range, simultaneously, from the UV to the near IR.One more disadvantage of Coude spectrographs concerning their design is that these instruments are hosted in a large room which ensures large inter-components distances needed to achieve high spectral resolution. The large room control of working parameters (temperature, humidity etc.) is a big challenge. 

A relevant solution for high-resolution spectral investigations challenges are fiber-fed echelle spectrographs. 
The efficiency of these instruments is twofold. First, their 2-D spectrum image is well suited to use effectively the full area of a 2-D CCD, thus offering the potential for the entire visible spectral range to be covered with a single exposure. And second, their design highly reduces the hosting space, thus improving spectrograph environment control. The first design of an echelle spectrograph for the 2-m telescope was a white pupil instrument, similar to FEROS (Kaufer et al., 1999) or to ELODIE (Baranne et al. 1996). Due to strong reduction of the preliminary planned budget for the spectrograph, we were pressed to develop a new design, different from the initial one, and based on commercial, off-the-shelf products, for all components where this was possible. Finally, our echelle spectrograph got a design similar to MUSICOS (Baudrand and Bohm, 1992) or to the University of Poznan spectrograph (Kami{\'n}ski et al., 2014).  

\section*{1. Spectrograph design and description of components}
\subsection*{1.1. Principle design}
\begin{figure}[!htb]
  \begin{center}
    \includegraphics[width=0.4\textwidth, trim = 0mm 0mm 0mm 0mm, clip=true]{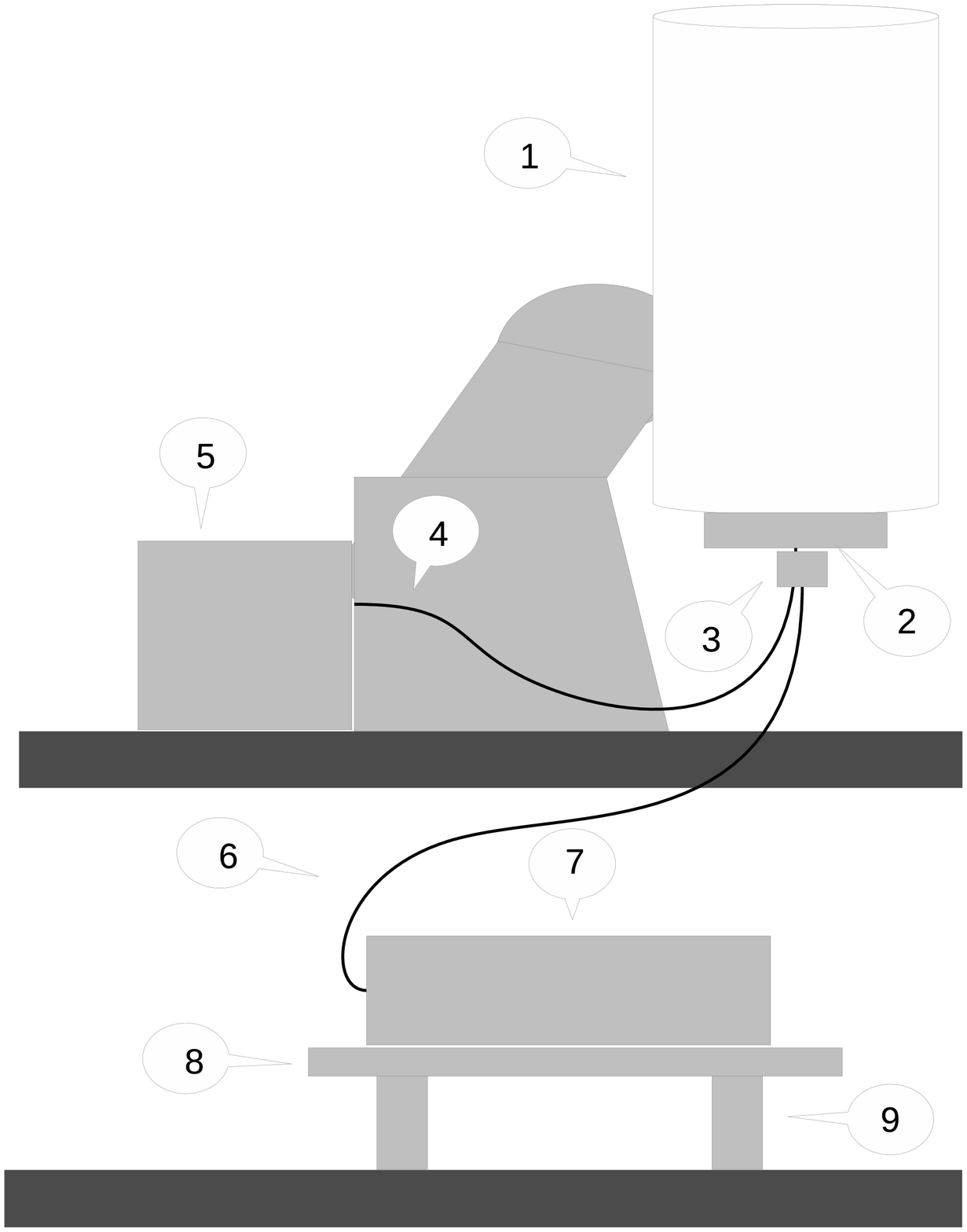}
    \label{fig:dome}
    \hfill
    \includegraphics[width=0.55\textwidth, trim = 0mm 50mm 0mm 50mm, clip=true]{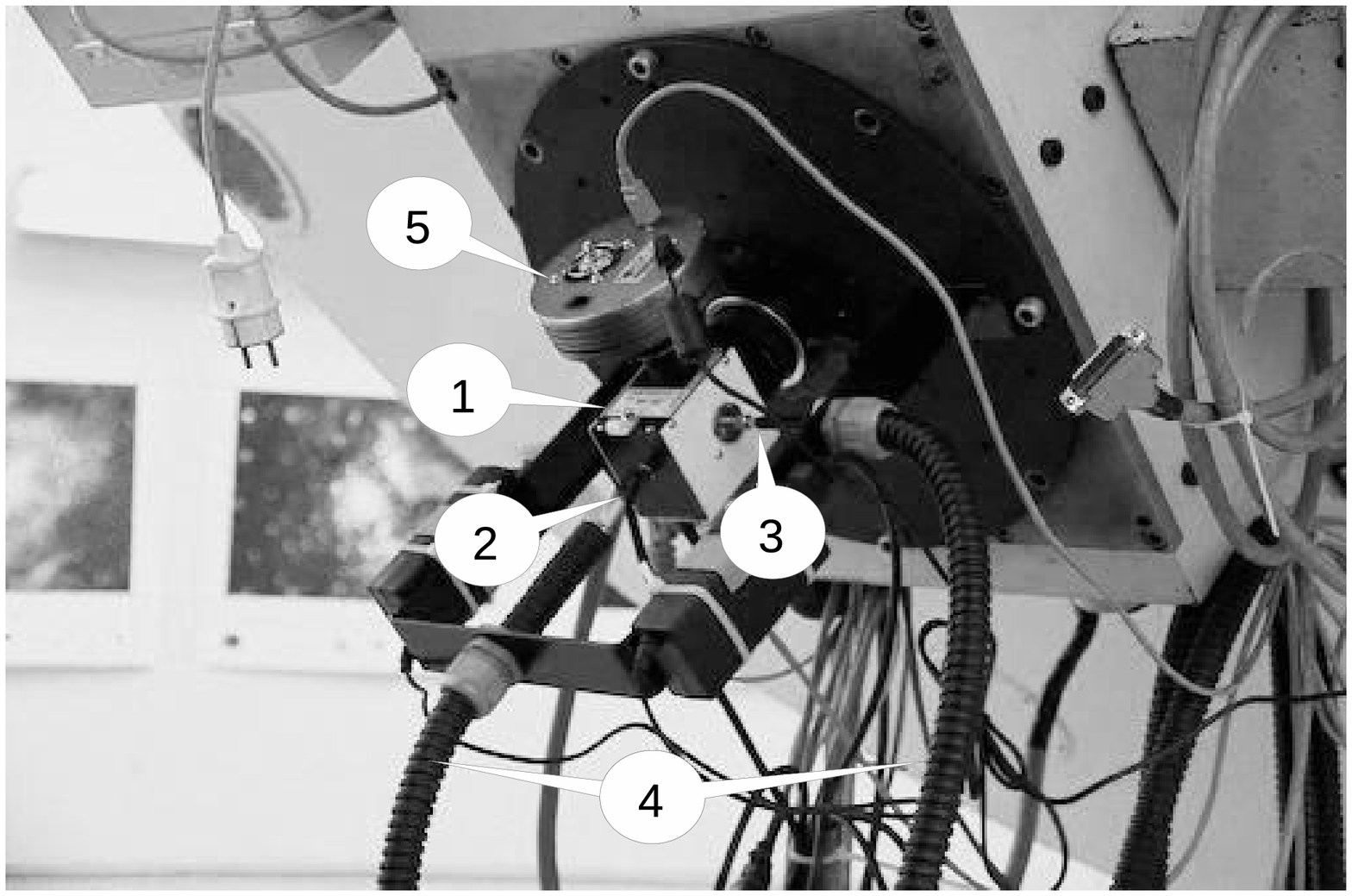}
    \caption[]{Left: ESpeRo principle layout, 1-telescope tube, 2 - RC-focus adapter, 3 - Fiber Injection and Guiding Unit, 4 - calibration fiber, 5 - calibration module, 6 - object fiber, 7 - spectrograph enclosure, 8 - optical table, 9 - pneumatic vibration isolators.  Right, components in the RC focal plane: 1- FIGU, 2-object fiber, 3-calibration fiber, 4-fiber protectors, 5-guiding CCD.}
  \end{center}
\end{figure}

ESpeRo is designed to be fiber-fed from the Ritchey-Chretien (RC) focus of the 2-m telescope. Figure \ref{fig:dome} shows the overall schema of the spectrograph design.
A Fiber Injection and Guiding Unit (FIGU) is mounted at the RC-focus.  The calibration unit, containing the calibration lamps (LED,tungsten and ThAr) is placed in the control room at the level of the telescope. 
The spectrograph itself is in a separate, thermally isolated room, one level below the telescope, next to the Coude spectrograph room.

\subsection*{1.2. Components in the RC-focus}

The light is transmitted from the RC focus to the spectrograph via an optical silica/silica multi-mode fiber with 50 $\mu$m core, 30 m length, and numerical aperture (N.A.) = 0.22.
The entrance of the fiber is mounted on the fiber injection and guiding unit (FIGU), produced by Shelyak instruments. Detailed description of FIGU can be found on the website of Shelyak Instruments\footnote{www.shelyak.com}. The main function of FIGU is to inject the focused light of the observed object into the object fiber. But this versatile unit is used, also, to transfer light from the calibration unit (for flat fielding (tungsten lamp and LED) and for wavelength calibration (ThAr lamp)) to the object fiber, and as well to transfer some amount of the object's light to a guiding CCD camera. The light beam from a star is focused by the telescope on a hole in the center of an inclined mirror. The central part of the image is transmitted through the hole, and the wings are reflected to the guiding camera. 
The calibration unit, comprising of the ThAr lamp for wavelength reference and the tungsten and LED for flat fieldig, is placed in a separate room at the level of the telescope. The calibration unit was also purchased from Shelyak. 

In order to achieve best match between the telescope focal ratio and the image scale (f/8 and 77.5 $\mu$m/\arcsec, respectively) with that of the FIGU, we purchased the f/9 version, with a 75 $\mu$m hole in the mirror. An optical system below the mirror, having transfer ratio 1.5:1, refocusses the 75 $\mu$m hole onto the tip of the 50 $\mu$m object fiber, and the focal ratio is changed from f/8 to f/5.3, which corresponds to a N.A. = 0.09, far below that of the fiber.      

The light beam from the calibration unit is transferred to the FIGU via a 200 $\mu$m fiber. An automatically controlled moving flat mirror reflects the beam into the hole of the mirror, and from there on the light follows the same optical path to the spectrograph as the beam from the object. 

The autoguiding is performed with an Atik Titan camera, which comprises of a CCD chip with 659 x 494 quadratic pixels of size 7.4 $\mu$m. Detailed description of the camera can be found on the vendor's webpage\footnote{http://www.atik-cameras.com/product/atik-titan/}. Before reaching the CCD, the reflected image of the object passes through a transfer optics with 1.5x magnification. Thus, on the guiding CCD, 1$\arcsec$ corresponds to 120 $\mu$m, the scale is 0.06$\arcsec/\mu$m, and the FOV = 40$\arcsec$ x 30$\arcsec$.

A dedicated mechanical mounting was manufactured to hold all the components in the RC-focus, with special attention to the stable fixation of both fibers, eliminating the mechanical load on their connectors at different positions of the telescope.    

\subsection*{1.3. The spectrograph}
\begin{figure}[!htb]
  \begin{center}
    \includegraphics[width=0.45\textwidth, trim = 0mm 0mm 0mm 0mm, clip=true]{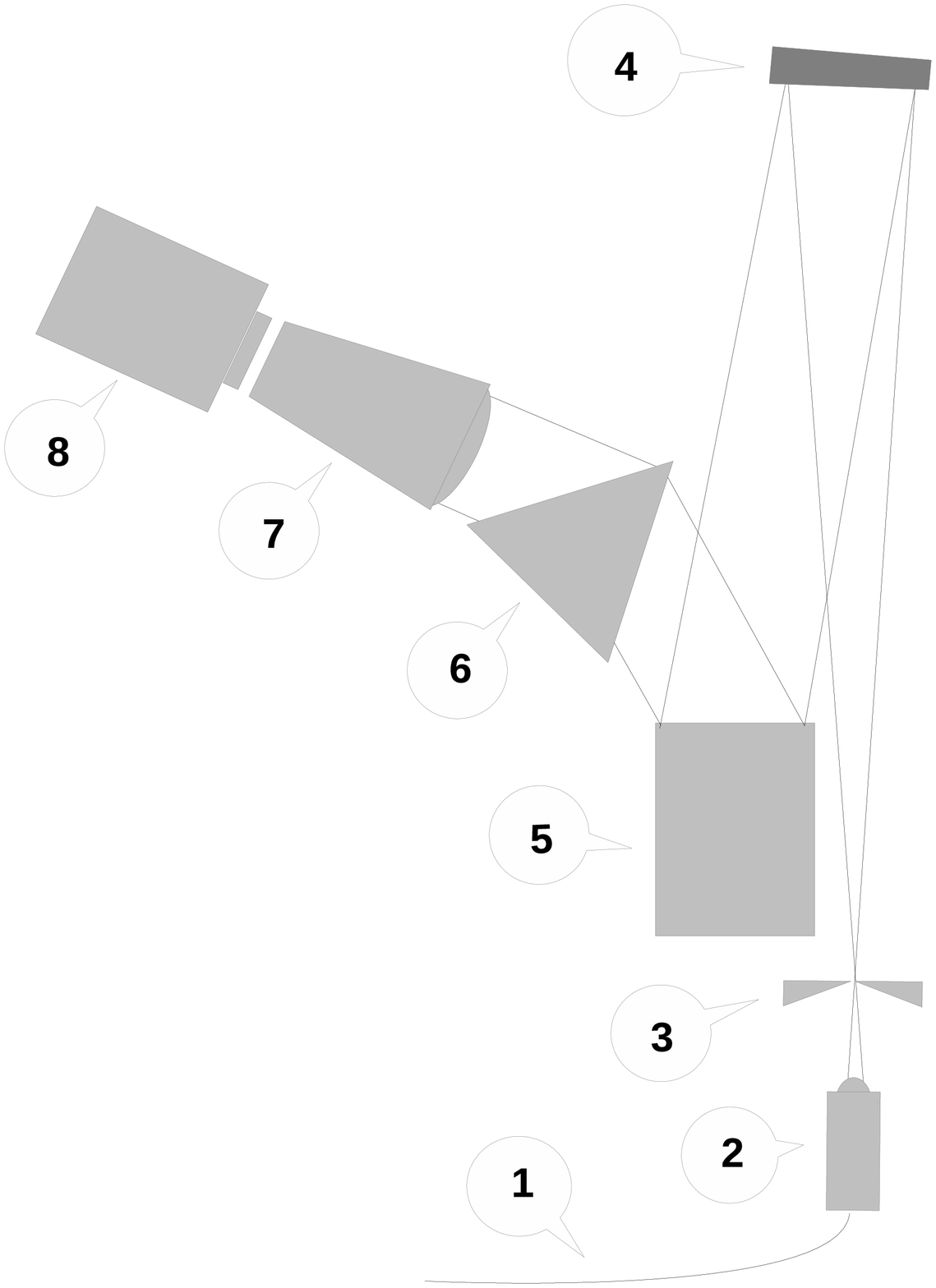}
    \hfill
    \includegraphics[width=0.45\textwidth, trim = 0mm 0mm 0mm 30mm, clip=true]{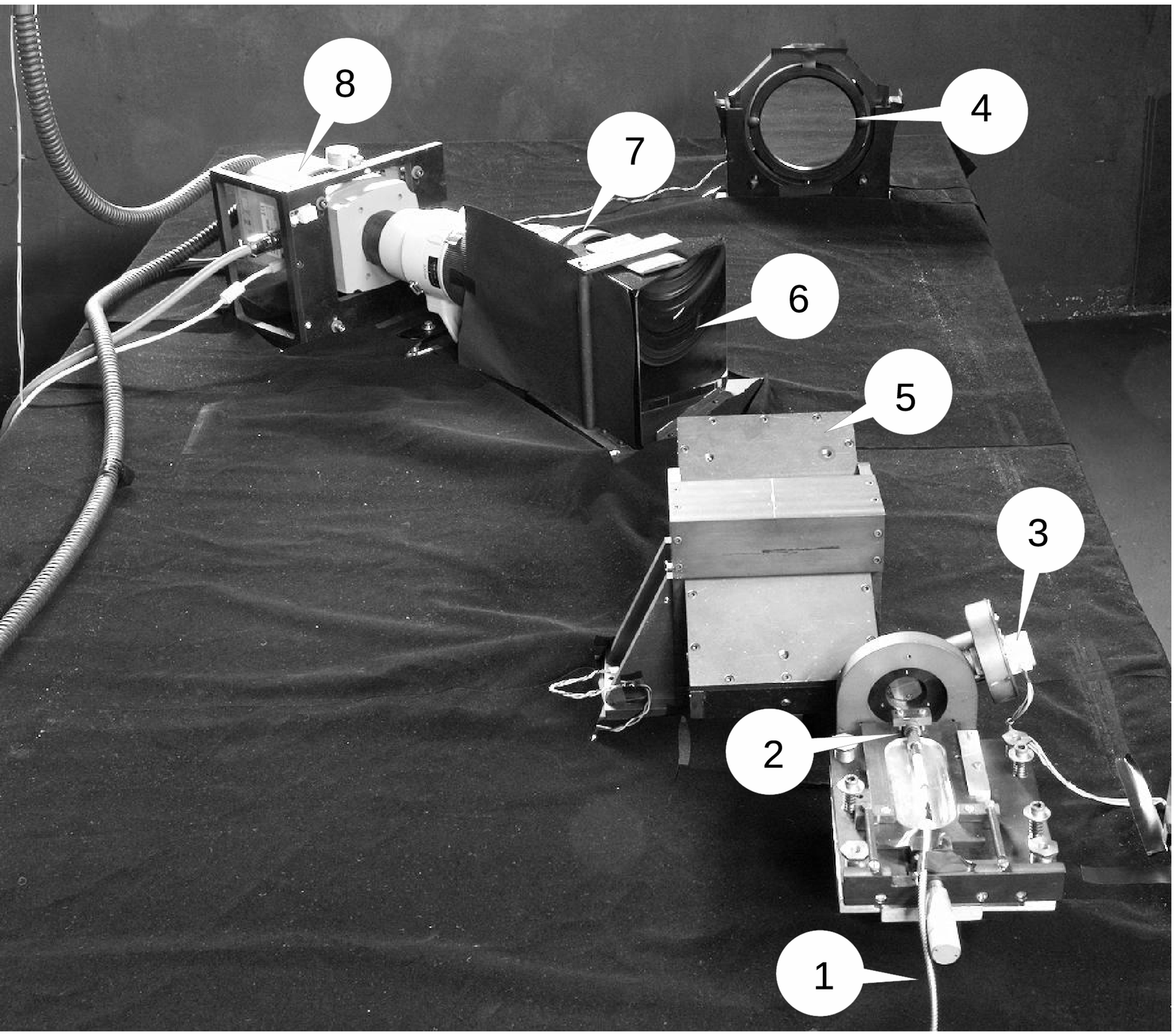}
    \caption[]{Left: ESpeRo principal opical layout , and Right: on-table view.}
    \label{fig:ontable}
  \end{center}
\end{figure}

Figure \ref{fig:ontable} shows the instrument in the spectrograph room. The fiber feeding of the spectrograph is marked with number 1. The exit of the fiber is mounted in the housing of an optical adapter (label 2 in fig. \ref{fig:ontable}, which serves for adjustment of the output beam to the f/15 beam of the collimator. This adaptation optics is just behind the slit (number 3 in fig. \ref{fig:ontable}). The output end of the object fiber with its optical adapter and the slit holder are firmly fixed on an originally designed and manufactured in NAO table with five orders of freedom. This mechanical device is aimed to ensure proper adjustment of incoming light beam onto the collimator focal plane.
The components 4,5,6,7, and 8 are the off-axis collimator (OAP), the echelle grating, the cross-dispersing prism, the camera lens, and the detector, respectively.    

The OAP is an off-the-shelf part purchased from Optical surfaces Ltd\footnote{www.optisurf.com}. 
It has a focal length of 1508.2 mm, an off-axis distance 103.9 mm, and a free aperture of 140 mm. 
Our optimized optical scheme is based on a parallel beam of 100 mm, thus we are working far from the edges of the mirror with an effective focal ratio of f/15. 

The R2 echelle grating has 37.5 lines/mm, its dimensions are 240x120x30 mm, and the blaze angle is 63.5$\arcdeg$. 
It works in a nearly Littrow regime, with an angle of 12$\arcdeg$ between the plane of dispersion and the directon to the crossdispersing prism.
The crossdisperser is a prism made of glass F108\footnote{lzos.ru/glass\_pdf/F8.pdf}, having characteristics similar to the F7 glass of Schott. 
The apex angle of the prism is 53.7$\arcdeg$ and the dimensions  of its entrance surface are 187x190 mm.

The imaging camera is a commercial Canon EF 400mm f/2.8L IS USM Lens.  
This objective lens has a diagonal FOV of 6$\arcdeg$10$\arcmin$, which is well adapted to the size of the detector (see next subsection). 

The detector is an iKon-L CCD camera, manufactured by Andor. It uses a  thermoelectric cooled,  back-illuminated, scientific-grade E2V CCD42-40 chip, comprising 2048x2048 quadratic pixels of size 13.5 $\mu$m. The camera is characterized by low readout noise, high sensitivity and high dynamic range. 
When reading with 50 kHz the gain is 1 e$^-$/ADU, and the RON is 3.4 electrons. The digitization is performed by a 16-bit analogue to digital converter, and the saturation level is over 88000 electrons.   

All components of the spectrograph are mounted on a honeycomb optical table with dimensions 2400 x 1200 mm, delivered by STANDA. The table itself stays on 4 pneumatic vibration isolators, which keep the level of the table and damp resonances caused by ambient vibrations. The vibration isolators are permanently connected to a cylinder of pressurized nitrogen.  

For better thermal isolation and stability and in order to keep  the spectrograph free of dust and other possible contaminants, it is enclosed in a non-hermetic box. All components of the spectrograph, except the CCD, are in the enclosure. This way any effect of the detector ventilation on the other parts of the spectrograph is avoided. To let the CCD outside, the box was manufactured with one hole between the end of the camera lens and the opening of the CCD.

\subsection*{1.4. Computer-controlled units}

The CCD detector is operated by an industrial PC running SOLIS, proprietary software supplied by Andor together with the camera. Because of the cable length limits of the USB connection, the PC is located relatively close to the camera, in the spectrograph room. 

Three additional single board computers (SBCs) operate (1) the setting of the slit width, (2) the monitoring of the temperature inside of the enclosure and in the spectrograph room environment, and (3) the autoguiding of the telescope during exposures.

The slit width is controlled by an Arduino UNO board\footnote{https://www.arduino.cc} located in the spectrograph room. This board is connected via RS-232 interface to the PC that operates the CCD camera, this way the slit width can be controlled remotely over the local area network (LAN) with a simple software tool developed for the purpose.

The temperature and humidity monitoring is done by a Raspberry Pi B SBC\footnote{http://www.raspberrypi.org}, which reads, records the sensor readings and provides web interface to the stored data. There are five sensors connected: three located in the spectrograph enclosure (on the CCD camera, on the collimator and on the grating) and two in the spectrograph room (on the ceiling and at the optical table). The monitoring of the temperature over several months in 2016 gives a mean temperature of 17.8$\pm$0.9$\arcdeg$C. The relative humidity shows larger fluctuations, between 30\% and 50\%. These values resulted from a very preliminary analysis of the monitoring data, and they give enough information about the future needs of better stabilisation of the spectrograph environment, especially of the humidity.

For the autoguiding another SBC is used, Raspberry Pi B2, which runs Lin\_guider\footnote{https://sourceforge.net/projects/linguider/}. Lin\_guider is a free software originally developed for amateur telescope guiding, but extended with new features which made it a professional-grade software. It is being actively developed and extended with new features to make it more suitable for guiding on the residual stellar image as the majority of the light goes into the object optical fibre and the guiding is performed only based on the tiny percentage of the side lobe light. Lin\_guider currently uses centroid algorithm, but tests are being performed with two new algorithms which will be part of the next major release. Both of them can be used for full frame guiding and all three have a subpixel precision. However each of them offers different benefits. One of them: DONUTS (McCormac et al.,2013) is suitabe for guiding on defocussed stellar image, useful for high precision photometry. And the second is specially designed for guiding using the side lobes of the stellar image, which is basically a stellar image with a moving hole in it, hence its name HOLES. The HOLES algorithm will be descibed in detail in a separate paper. 

The guiding camera, Atik Titan, is equipped with standard ST4 guiding port. For that purpose a special adapter is created to make it possible to control the 2\_meter telescope via the ST4 port. This approach has another benefit  - any of the wide range of cameras, supported by Lin\_guider, that is equipped with guiding port, can be used instead.
 
The detector PC and the three SBCs are controlled remotely over the LAN. Thus, entering the spectrograph room is necessary only on exceptional occasions. 
    
\section*{2. Simulated distribution of orders}
For calculation of the orders and their corresponding wavelengths we used the grating equation, as given by Schroeder and Hilliard (1980):
\begin{equation}
(m\lambda)/\sigma = \cos\gamma(\sin\alpha + \sin\beta),
\end{equation}

\noindent where $m$ is the order of the wavelength $\lambda$ which is diffracted from an echelle grating of grove spacing $\sigma$, and 
$\alpha$, $\beta$, and $\gamma$ are the angles of incidence, diffraction and deviation from the plane defined by the incident and diffracted rays, respectively. 

\begin{table}[!htb]
\begin{center}
  \caption{Calculated orders. $\lambda_u$, $\lambda_c$, and $\lambda_d$  refer to the up, central, and down wavelength in each order, respectively. This designation corresponds to the order distribution presented in Fig. \ref{fig:orders}. FSR is acronym for Free Spectral Range. For more details see the text.}
  \setlength{\tabcolsep}{2.55mm}
  \begin{tabular}{cccccccc}
Order &$\lambda_u$&$\lambda_c$&$\lambda_d$&$\lambda_d$-$\lambda_u$&FSR&$\lambda_d$-$\lambda_u$-FSR&Dispersion\\     
      &   \AA     &   \AA     &    \AA    &     \AA               &\AA&       \AA                 &    \AA/px  \\ 
      \hline\\
  53  &  8849.97 &   8927.87  &   9000.66  &   150.69  &   168.45 &    -17.76  &   0.0736 \\ 
  54  &  8686.08 &   8762.54  &   8833.98  &   147.90  &   162.27 &    -14.37  &   0.0722 \\ 
  55  &  8528.15 &   8603.22  &   8673.36  &   145.21  &   156.42 &    -11.21  &   0.0709 \\ 
  56  &  8375.86 &   8449.59  &   8518.48  &  142.62   &  150.89  &     -8.27  &   0.0696 \\ 
  57  &  8228.92 &   8301.36  &   8369.03  &   140.11  &   145.64 &     -5.52  &   0.0684 \\ 
  58  &  8087.04 &   8158.23  &   8224.74  &   137.70  &   140.66 &     -2.96  &   0.0672 \\ 
  59  &  7949.97 &   8019.95  &   8085.34  &   135.36  &   135.93 &     -0.57  &   0.0661 \\ 
  60  &  7817.47 &   7886.29  &   7950.58  &   133.11  &   131.44 &      1.67  &   0.0650 \\ 
  61  &  7689.32 &   7757.00  &   7820.24  &   130.93  &   127.16 &      3.76  &   0.0639 \\ 
  62  &  7565.30 &   7631.89  &   7694.11  &   128.81  &   123.10 &      5.72  &   0.0629 \\ 
  63  &  7445.21 &   7510.75  &   7571.98  &   126.77  &   119.22 &      7.55  &   0.0619 \\ 
  64  &  7328.88 &   7393.40  &   7453.67  &   124.79  &   115.52 &      9.27  &   0.0609 \\ 
  65  &  7216.13 &   7279.65  &   7339.00  &   122.87  &   111.99 &     10.88  &   0.0600 \\ 
  66  &  7106.79 &   7169.35  &   7227.80  &   121.01  &   108.63 &     12.38  &   0.0591 \\ 
  67  &  7000.72 &   7062.35  &   7119.92  &   119.20  &   105.41 &     13.79  &   0.0582 \\ 
  68  &  6897.77 &   6958.49  &   7015.22  &   117.45  &   102.33 &     15.12  &   0.0573 \\ 
  69  &  6797.80 &   6857.64  &   6913.55  &   115.75  &    99.39 &     16.36  &   0.0565 \\ 
  70  &  6700.69 &   6759.68  &   6814.78  &   114.09  &    96.57 &     17.53  &   0.0557 \\ 
  71  &  6606.32 &   6664.47  &   6718.80  &   112.49  &    93.87 &     18.62  &   0.0549 \\ 
  72  &  6514.56 &   6571.91  &   6625.49  &   110.92  &    91.28 &     19.65  &   0.0542 \\ 
  73  &  6425.32 &   6481.88  &   6534.73  &   109.40  &    88.79 &     20.61  &   0.0534 \\ 
  74  &  6338.49 &   6394.29  &   6446.42  &   107.93  &    86.41 &     21.52  &   0.0527 \\ 
  75  &  6253.98 &   6309.03  &   6360.47  &   106.49  &    84.12 &     22.37  &   0.0520 \\ 
  76  &  6171.69 &   6226.02  &   6276.78  &   105.09  &    81.92 &     23.16  &   0.0513 \\ 
  77  &  6091.54 &   6145.16  &   6195.26  &   103.72  &    79.81 &     23.91  &   0.0506 \\ 
  78  &  6013.44 &   6066.38  &   6115.83  &   102.39  &    77.77 &     24.62  &   0.0500 \\ 
  79  &  5937.32 &   5989.59  &   6038.42  &   101.10  &    75.82 &     25.28  &   0.0494 \\ 
  80  &  5863.10 &   5914.72  &   5962.94  &    99.83  &    73.93 &     25.90  &   0.0487 \\ 
  81  &  5790.72 &   5841.69  &   5889.32  &    98.60  &    72.12 &     26.48  &   0.0481 \\ 
  82  &  5720.10 &   5770.46  &   5817.50  &    97.40  &    70.37 &     27.03  &   0.0476 \\ 
  83  &  5651.19 &   5700.93  &   5747.41  &    96.22  &    68.69 &     27.54  &   0.0470 \\ 
  84  &  5583.91 &   5633.06  &   5678.99  &    95.08  &    67.06 &     28.02  &   0.0464 \\ 
  85  &  5518.22 &   5566.79  &   5612.18  &    93.96  &    65.49 &     28.47  &   0.0459 \\ 
  86  &  5454.05 &   5502.06  &   5546.92  &    92.87  &    63.98 &     28.89  &   0.0453 \\ 
  87  &  5391.36 &   5438.82  &   5483.16  &    91.80  &    62.52 &     29.28  &   0.0448 \\ 
  88  &  5330.10 &   5377.01  &   5420.85  &    90.76  &    61.10 &     29.65  &   0.0443 \\ 
  89  &  5270.21 &   5316.60  &   5359.94  &    89.74  &    59.74 &     30.00  &   0.0438 \\ 
  90  &  5211.65 &   5257.53  &   5300.39  &    88.74  &    58.42 &     30.32  &   0.0433 \\ 
  91  &  5154.38 &   5199.75  &   5242.14  &    87.76  &    57.14 &     30.62  &   0.0429 \\ 
  92  &  5098.35 &   5143.23  &   5185.16  &    86.81  &    55.90 &     30.90  &   0.0424 \\ 
  93  &  5043.53 &   5087.93  &   5129.41  &    85.88  &    54.71 &     31.17  &   0.0419 \\ 
  94  &  4989.88 &   5033.80  &   5074.84  &    84.96  &    53.55 &     31.41  &   0.0415 \\ 
  95  &  4937.35 &   4980.81  &   5021.42  &    84.07  &    52.43 &     31.64  &   0.0410 \\ 
  96  &  4885.92 &   4928.93  &   4969.11  &    83.19  &    51.34 &     31.85  &   0.0406 \\ 
  97  &  4835.55 &   4878.12  &   4917.89  &    82.34  &    50.29 &     32.05  &   0.0402 \\ 
  98  &  4786.21 &   4828.34  &   4867.70  &    81.50  &    49.27 &     32.23  &   0.0398 \\ 
  99  &  4737.86 &   4779.57  &   4818.53  &    80.67  &    48.28 &     32.39  &   0.0394 \\ 
 100  &  4690.48 &   4731.77  &   4770.35  &    79.87  &    47.32 &     32.55  &   0.0390 \\ 
 101  &  4644.04 &   4684.92  &   4723.12  &    79.07  &    46.39 &     32.69  &   0.0386 \\ 
 102  &  4598.51 &   4638.99  &   4676.81  &    78.30  &    45.48 &     32.82  &   0.0382 \\ 
 103  &  4553.87 &   4593.95  &   4631.41  &    77.54  &    44.60 &     32.94  &   0.0379 \\ 
 104  &  4510.08 &   4549.78  &   4586.87  &    76.79  &    43.75 &     33.05  &   0.0375 \\ 
 105  &  4467.13 &   4506.45  &   4543.19  &    76.06  &    42.92 &     33.14  &   0.0371 \\ 
 106  &  4424.98 &   4463.94  &   4500.33  &    75.34  &    42.11 &     33.23  &   0.0368 \\ 
 107  &  4383.63 &   4422.22  &   4458.27  &    74.64  &    41.33 &     33.31  &   0.0364 \\ 
 108  &  4343.04 &   4381.27  &   4416.99  &    73.95  &    40.57 &     33.38  &   0.0361 \\ 
 109  &  4303.20 &   4341.08  &   4376.47  &    73.27  &    39.83 &     33.44  &   0.0358 \\ 
 110  &  4264.08 &   4301.61  &   4336.68  &    72.60  &    39.11 &     33.50  &   0.0355 \\ 
 111  &  4225.66 &   4262.86  &   4297.61  &    71.95  &    38.40 &     33.55  &   0.0351 \\ 
 112  &  4187.93 &   4224.80  &   4259.24  &    71.31  &    37.72 &     33.59  &   0.0348 \\ 
 113  &  4150.87 &   4187.41  &   4221.55  &    70.68  &    37.06 &     33.62  &   0.0345 \\ 
 114  &  4114.46 &   4150.68  &   4184.52  &    70.06  &    36.41 &     33.65  &   0.0342 \\ 
 115  &  4078.68 &   4114.58  &   4148.13  &    69.45  &    35.78 &     33.67  &   0.0339 \\ 
 116  &  4043.52 &   4079.11  &   4112.37  &    68.85  &    35.16 &     33.68  &   0.0336 \\ 
 117  &  4008.96 &   4044.25  &   4077.22  &    68.26  &    34.57 &     33.69  &   0.0333 \\ 
 118  &  3974.99 &   4009.98  &   4042.67  &    67.68  &    33.98 &     33.70  &   0.0330 \\ 
 119  &  3941.58 &   3976.28  &   4008.70  &    67.11  &    33.41 &     33.70  &   0.0328 \\ 
 120  &  3908.74 &   3943.14  &   3975.29  &    66.55  &    32.86 &     33.69  &   0.0325 \\ 
\end{tabular}
\label{tab:orders}
\end{center}
\end{table} 

\begin{figure}[!htb]
  \begin{center}
    \includegraphics[width=0.8\textwidth, trim = 0mm 0mm 0mm 20mm, clip=true]{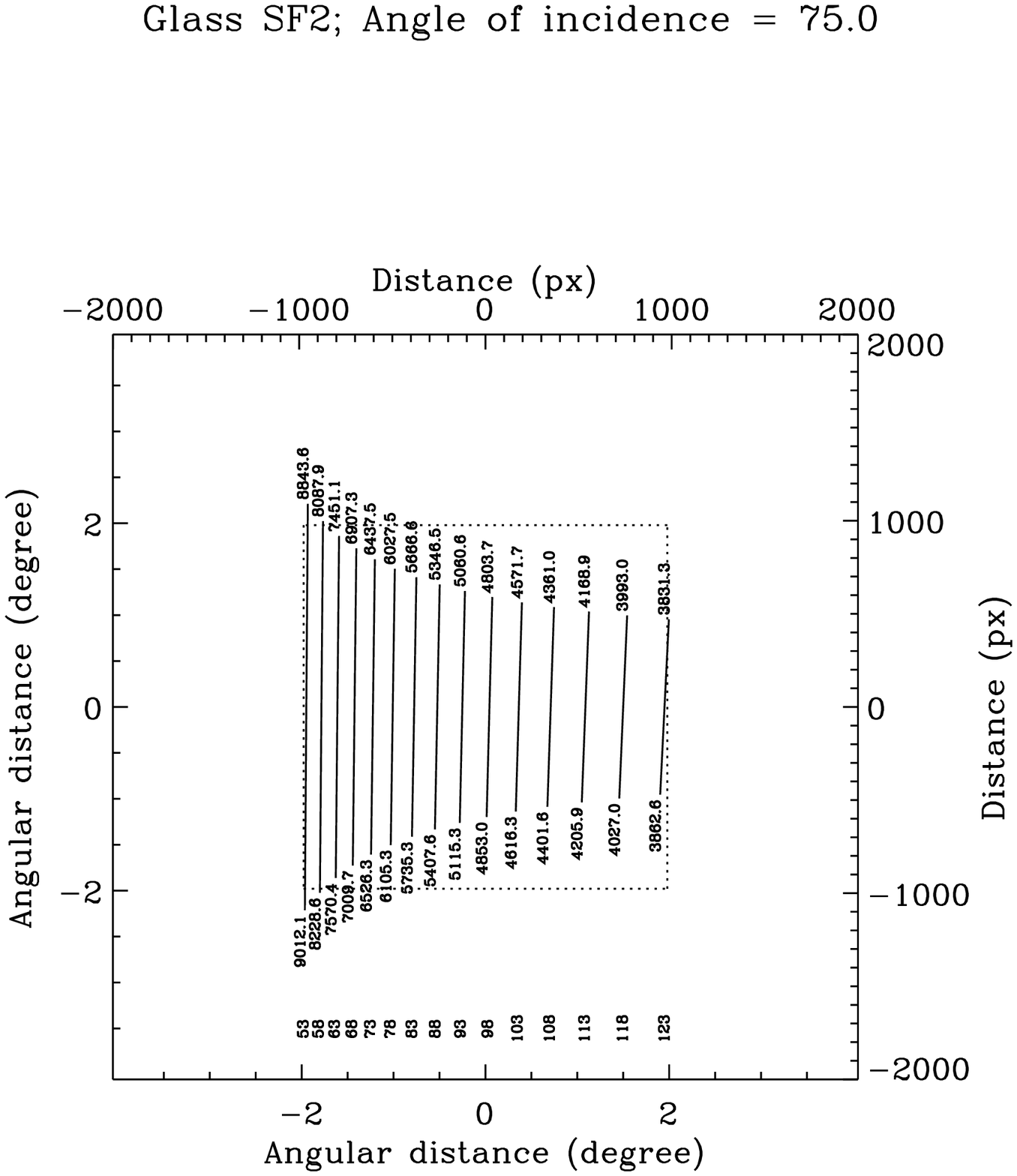}
    \caption[]{Distribution of orders on the detector.}
    \label{fig:orders}
  \end{center}
\end{figure}

Table \ref{tab:orders} lists the distribution of wavelengths among the different orders. A total of 71 orders cover the spectral range from 3813 {\AA} to 9000 {\AA}.  The titles of the columns in that table explain their content. In the sixth column, the negative values in the first seven rows are the gaps in the corresponding orders. The separation of the orders over the CCD chip is achieved with putting the cross-disperser in the beam of diffracted light. Figure \ref{fig:orders} shows the distribution of the orders over the detector. For clarity, every fifth order is given, and marked with the order number and its wavelength limits.
Evidently the CCD is effectively used, the orders are spread over the full frame of the chip. 

\section*{3 Performance of ESpeRo}

\subsection*{3.1 Identified orders}
\label{subsec:identi_ord}
The extraction of all orders in the obtained spectra requires a well exposed set of flat fields, spectra of hot stars or solar spectra, giving sufficient signal over the whole range of wavelengths. The extracted orders are listed in Tab.~\ref{tab:apertures}. The central wavelength and dispersion  were derived from wavelength calibration with the ThAr lines, identified in every order.  The separation between the extracted extremely blue orders amounts to 47 pixels and between the extremely red orders - 17 pixels.
    
\begin{table}[h!tb]
\begin{center}
  \caption{Extracted orders}
  \setlength{\tabcolsep}{5 mm}
  \begin{tabular}{ccccc}
aperture &  order  & row center    &   $\lambda_c$   & order dispersion\\
number   &  number & pixel         & {\AA}           & {\AA}/px   \\
      \hline\\
  2  &    53   &      66.4    &    8939.8   &    0.0771  \\
  3  &    54   &      83.1    &    8774.3   &    0.0756  \\
  4  &    55   &      99.8    &    8614.8   &    0.0743  \\
  5  &    56   &     116.7    &    8460.9   &    0.0729  \\
  6  &    57   &     133.8    &    8312.5   &    0.0717  \\
  7  &    58   &     151.0    &    8169.2   &    0.0704  \\
  8  &    59   &     168.5    &    8030.7   &    0.0693  \\
  9  &    60   &     186.2    &    7896.9   &    0.0681  \\
 10  &    61   &     204.1    &    7767.4   &    0.0670  \\
 11  &    62   &     222.2    &    7642.2   &    0.0659  \\
 12  &    63   &     240.7    &    7520.9   &    0.0649  \\
 13  &    64   &     259.3    &    7403.4   &    0.0639  \\
 14  &    65   &     278.2    &    7289.5   &    0.0629  \\
 15  &    66   &     297.3    &    7179.0   &    0.0619  \\
 16  &    67   &     316.8    &    7071.9   &    0.0610  \\
 17  &    68   &     336.5    &    6967.9   &    0.0601  \\
 18  &    69   &     356.5    &    6866.9   &    0.0592  \\
 19  &    70   &     376.7    &    6768.8   &    0.0584  \\
 20  &    71   &     397.2    &    6673.5   &    0.0576  \\
 21  &    72   &     418.1    &    6580.8   &    0.0568  \\
 22  &    73   &     439.3    &    6490.7   &    0.0560  \\
 23  &    74   &     460.8    &    6403.0   &    0.0553  \\
 24  &    75   &     482.6    &    6317.6   &    0.0545  \\
 25  &    76   &     504.8    &    6234.5   &    0.0538  \\
 26  &    77   &     527.3    &    6153.5   &    0.0531  \\
 27  &    78   &     550.1    &    6074.6   &    0.0524  \\
 28  &    79   &     573.3    &    5997.8   &    0.0518  \\
 29  &    80   &     596.9    &    5922.8   &    0.0511  \\
 30  &    81   &     620.8    &    5849.7   &    0.0505  \\
 31  &    82   &     645.1    &    5778.3   &    0.0499  \\
 32  &    83   &     669.8    &    5708.7   &    0.0493  \\
 33  &    84   &     694.9    &    5640.8   &    0.0487  \\
 34  &    85   &     720.2    &    5574.4   &    0.0481  \\
 35  &    86   &     746.1    &    5509.6   &    0.0476  \\
 36  &    87   &     772.3    &    5446.3   &    0.0470  \\
 37  &    88   &     799.0    &    5384.4   &    0.0465  \\
 38  &    89   &     826.1    &    5323.9   &    0.0460  \\
 39  &    90   &     853.7    &    5264.8   &    0.0455  \\
 40  &    91   &     881.7    &    5206.9   &    0.0450  \\
 41  &    92   &     910.1    &    5150.3   &    0.0445  \\
 42  &    93   &     939.0    &    5095.0   &    0.0440  \\
 43  &    94   &     968.2    &    5040.8   &    0.0435  \\
 44  &    95   &     998.1    &    4987.7   &    0.0431  \\
 45  &    96   &    1028.4    &    4935.8   &    0.0426  \\
 46  &    97   &    1059.2    &    4884.9   &    0.0422  \\
 47  &    98   &    1090.6    &    4835.0   &    0.0417  \\
 48  &    99   &    1122.3    &    4786.2   &    0.0413  \\
 49  &   100   &    1154.8    &    4738.4   &    0.0409  \\
 50  &   101   &    1187.7    &    4691.5   &    0.0405  \\
 51  &   102   &    1221.1    &    4645.5   &    0.0401  \\
 52  &   103   &    1255.1    &    4600.4   &    0.0397  \\
 53  &   104   &    1289.7    &    4556.1   &    0.0393  \\
 54  &   105   &    1324.9    &    4512.8   &    0.0389  \\
 55  &   106   &    1360.7    &    4470.2   &    0.0386  \\
 56  &   107   &    1397.0    &    4428.4   &    0.0382  \\
 57  &   108   &    1434.1    &    4387.4   &    0.0379  \\
 58  &   109   &    1471.8    &    4347.2   &    0.0375  \\
 59  &   110   &    1510.1    &    4307.7   &    0.0372  \\
 60  &   111   &    1549.1    &    4268.9   &    0.0368  \\
 61  &   112   &    1588.9    &    4230.8   &    0.0365  \\
 62  &   113   &    1629.2    &    4193.3   &    0.0362  \\
 63  &   114   &    1670.5    &    4156.6   &    0.0358  \\
 64  &   115   &    1712.4    &    4120.4   &    0.0355  \\
 65  &   116   &    1755.1    &    4084.9   &    0.0352  \\
 66  &   117   &    1798.8    &    4050.0   &    0.0349  \\
 67  &   118   &    1843.9    &    4015.7   &    0.0346  \\
 68  &   119   &    1886.6    &    3982.0   &    0.0343  \\
 69  &   120   &    1934.0    &    3948.8   &    0.0340  \\

\end{tabular}
\label{tab:apertures}
\end{center}
\end{table}

\begin{figure}[!htb]
  \begin{center}
    \includegraphics[width=0.7\textwidth, trim = 0mm 0mm 0mm 0mm, clip=true]
    {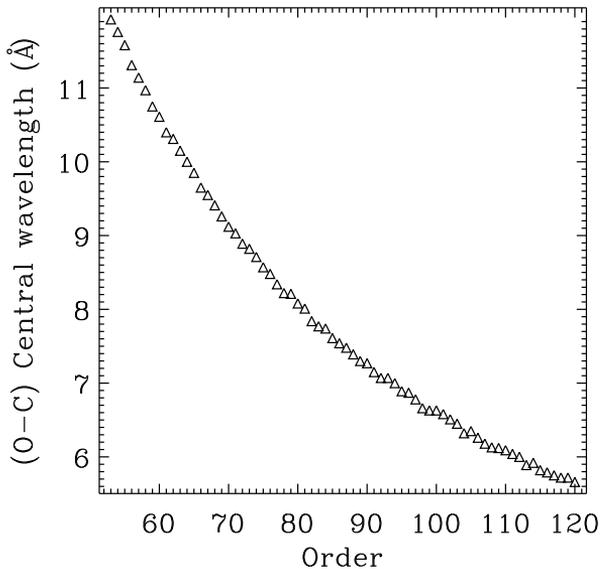}
    \caption[]{Difference between the observed and calculated central wavelengths of the orders.}
    \label{fig:O_C}
  \end{center}
\end{figure}

Comparison of the central wavelengths of the extracted orders with the calculated ones shows  small differences which are gradually decreasing from about 12 {\AA} in order 53 (8940 {\AA}) to below 6 {\AA} in order 120 (3950 {\AA}). This dependence is presented in Fig. \ref{fig:O_C}. The relative values of these differences are almost constant over the entire spectral range. 

\begin{figure}[!htb]
  \begin{center}
    \includegraphics[width=0.7\textwidth, trim = 0mm 0mm 0mm 0mm, clip=true]{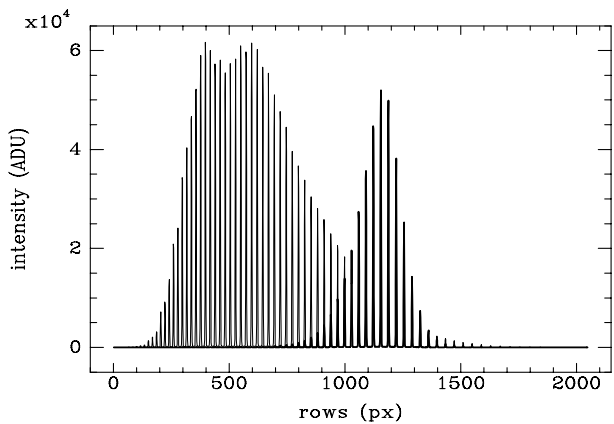}
    \caption[]{Tungsten (left peaked) and LED (right peaked) light distribution across the orders. The tungsten lamp peaks in the interval 5400-7200 {\AA}. The LED lamp peaks at $\sim$ 4800 {\AA}.}
    \label{fig:tung-led}
  \end{center}
\end{figure}

The role of high quality flat fields is very important in the process of extraction of orders. 
The flat field must not only correct for the pixel-to-pixel sensitivity, but it has also to compensate the blaze of the echelle grating and remove the fringing in the red spectral region. In order to fulfil all these functions at the same time it is necessary to have sufficient signal over the entire spectral range. It is hard to meet this requirement, especially for the blue orders where the signal of the flat field is very low.   This situation is well illustrated on Fig. \ref{fig:tung-led}, where a profile of a flat field, taken in the cross-dispersion direction, is shown. The blue end is at the right, and it is seen that the intensity of the LED drops very fast in that direction. The decreasing quantum efficiency of the CCD in the blue spectral region additionally contributes for lower signal in that region. In order to obtain sufficient signal in the last blue orders it is necessary to make  flat fields with longer exposures, and subsequently to combine the flats with different exposures into an usable superflat.    

Another feature of the LED and tungsten lamp is that their spectra, considered in a given single order, are inconsistent. Probably, the reason is the different color temperature of the tungsten and the LED. Up to date it is not clear which of the lamps better removes the echelle blaze function, therefore we recommend to expose the tungsten and LED lamps separately, and find empirically which lamp gives better results for every particular order.

\begin{figure}[!htb]
  \begin{center}
    \includegraphics[width=0.7\textwidth, trim = 0mm 0mm 0mm 0mm, clip=true]{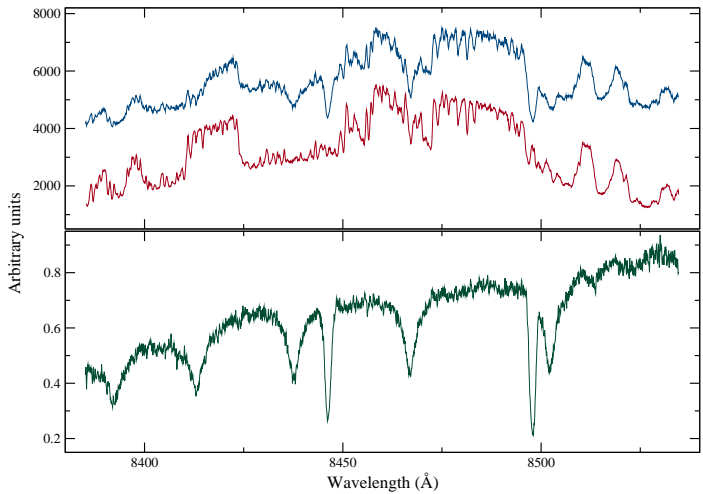}
    \caption[]{Elimination of the strong fringes in the red spectral region after flat-field correction. Upper panel: part of the spectrum of  $\varepsilon$~Aur (upper curve). Here the lower curve is the 
    flat field in the same spectral region. Lower panel: The spectrum of the star after division by the flat field (for more information see the text).}
    \label{fig:fringes}
  \end{center}
\end{figure}

As mentioned above the extraction of orders depends on high-quality flat fields. This dependence is especially critical in the red spectral region, where fringe patterns start to appear close to H$\alpha$ and are getting stronger with increasing wavelength. For wavelengths greater than 7000 {\AA} the fringing is so strong that it can compromise the scientific information contained in the spectra, if they are not   processed correctly with suitable flat fields. An example of removal of fringes in the range 8380 {\AA} - 8530 {\AA} is presented in Fig. \ref{fig:fringes}. The final result, presented in the lower panel of that figure, shows that after the division by the flat field the fringe patterns are excellently removed, and the blaze function of the grating and the influence of the pixel-to-pixel sensitivity are eliminated simultaneously.

\subsection*{3.2 Resolving power}

\begin{figure}[!htb]
  \begin{center}
    \includegraphics[width=\textwidth, trim = 0mm 0mm 0mm 0mm, clip=true]{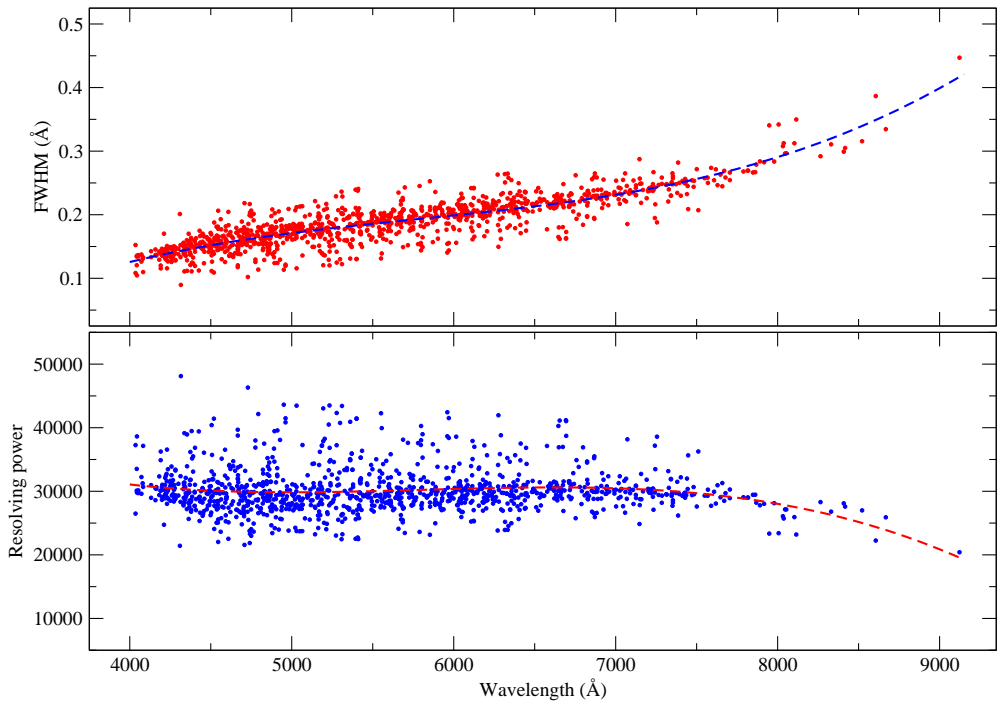}
    \caption[]{Upper panel: Full width at half maximum of 1178 lines, measured with fully opened slit. Lower panel: Resolving power, calculated with the FWHM presented in the upper panel.  
   }. 
    \label{fig:noslit_resolution}
  \end{center}
\end{figure}

One example of the derived resolving power of the spectrograph (R = $\lambda$/$\delta\lambda$) is presented in Fig. \ref{fig:noslit_resolution}. 
The Full Width at Half Maximum (FWHM =$\delta\lambda$) of more than thousand lines distributed over all orders were measured by fitting the lines with Gaussian function.  Some of the lines, especially in the blue region, appear in adjacent orders. These lines are measured twice, but the contribution of every measurement has the same weight, as in these cases the measurements are independent. 

\begin{figure}[!htb]
  \begin{center}
    \includegraphics[width=0.75\textwidth, trim = 0mm 0mm 0mm 0mm, clip=true]
    {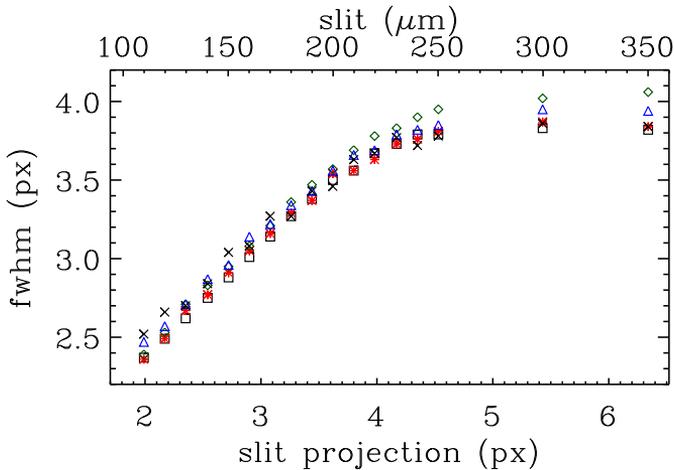}
    \caption[]{Measured Gaussian approximation FWHM of a set of lines, distributed in five orders, vs. projected slit width.  The upper abscissa shows the slit width in micrometers. As they are indistinguishable on the figure, the uncertainties of the FWHM values can be seen in column (7) of Tab. \ref{tab:losses}. }
    \label{fig:slit-fwhm}
  \end{center}
\end{figure}

There is a possibility to obtain spectra with higher resolutions by decreasing the width of the slit, placed in the OAP collimator focal plane. Of course, going to smaller widths, after a given value, a decrease of the signal will be experienced. In order to estimate the expected losses, the FWHM of calibration lines, selected within five randomly distributed orders, were measured. The result of these measurements are shown in 
Fig. \ref{fig:slit-fwhm}. Here, the upper abscissa shows the width at which the slit is set, and the lower one is its projection on the chip. The projection is  calculated by the relation describing the transfer through the spectrograph optics ($w/(k\times s)$), where 
{\it w} is the slit width in microns, $s$ is the CCD pixel size($13.5\mu m$), and {\it k} is the spectrograph transmission factor ($k = gm\times Fcol/Fcam$, where {\it gm} is the grating magnification factor, {\it Fcol} and {\it Fcam} are the OAP  and  CANON focal lengths, respectively).  The most remarkable feature in this figure is the constant FWHM for slit widths greater than 200 $\mu$m, which is an indication that over this limit the spectra are obtained with the maximum possible illumination. 

\begin{figure}[!htb]
  \begin{center}
    \includegraphics[width=0.75\textwidth, trim = 0mm 30mm 0mm 30mm, clip=true]{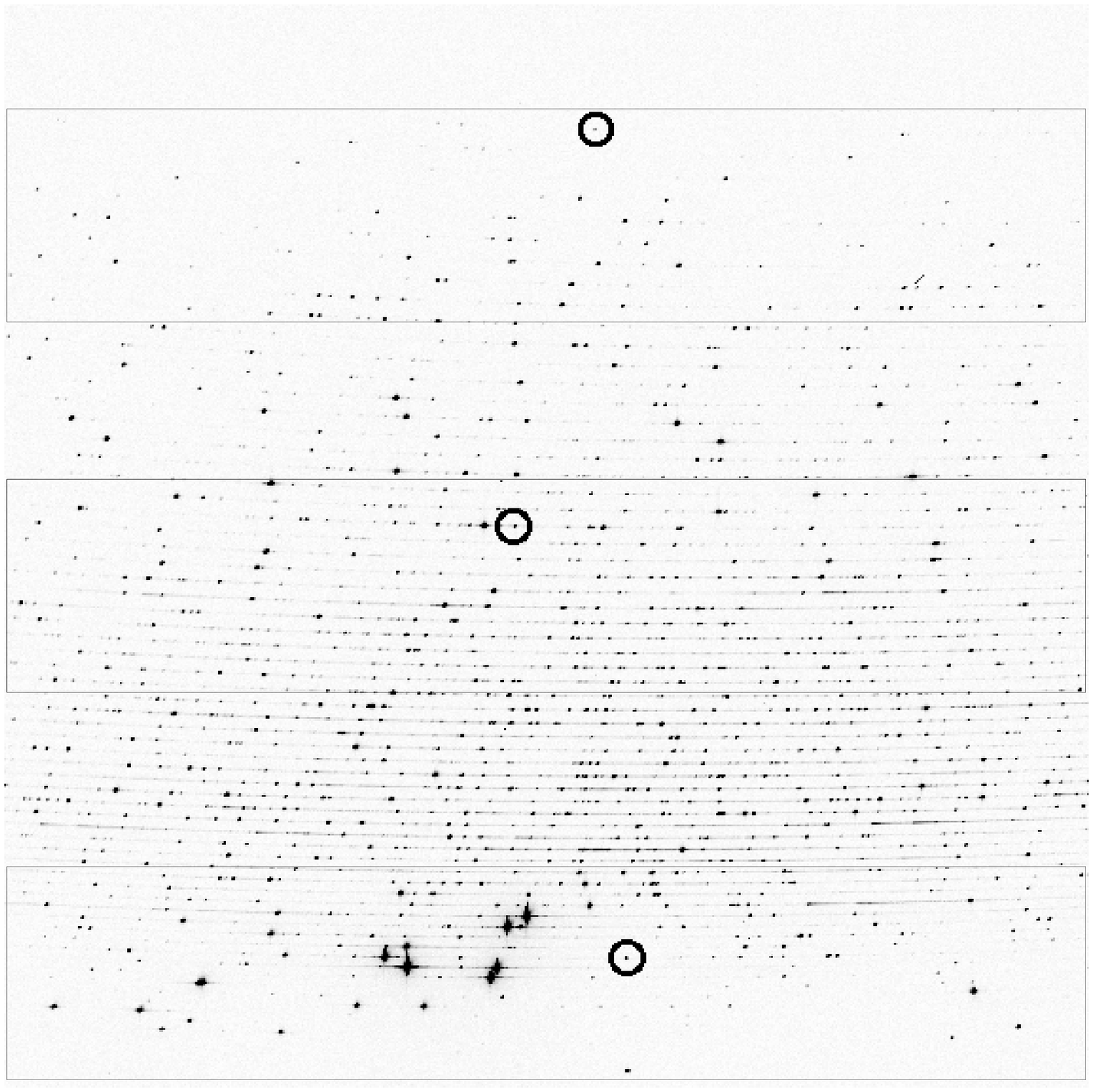}    
    \caption[]{CCD regions where ThAr lines full widths were measured. In the text these regions are referred to as blue(up), green(center) and red(down). The circles mark the three lines selected for the analysis presented in Fig. \ref{fig:slit_project} and Fig. \ref{fig:resolve_power}.}   
    \label{fig:ccd_rgb}
    \includegraphics[width=0.75\textwidth, trim = 0mm 0mm 0mm -10mm, clip=true]{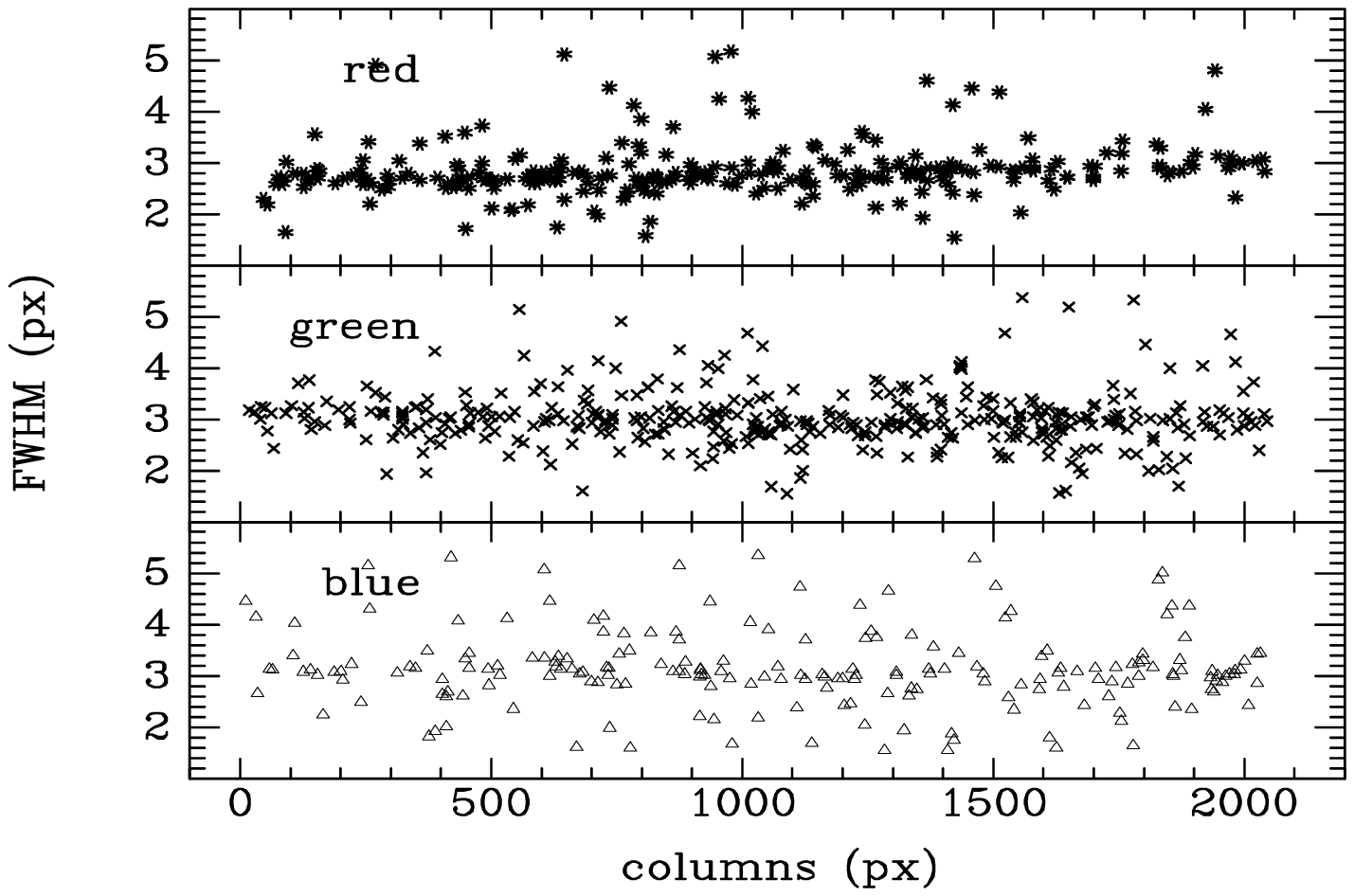}
    \caption[]{FWHM along dispersion at the three spectral regions marked on Fig. \ref{fig:ccd_rgb}. The most remarkable feature is the stronger scatter in the blue spectral region, caused by the lower CCD efficiency in the blue range.}
    \label{fig:fwhm_cols}
  \end{center}
\end{figure}

The relatively uniform distribution of R over the whole spectral range was achieved after careful focusing, including changes in the distance and the tilt between the camera lens and the detector.  To check the uniformity of the optical alignment we measured the full widths of many ThAr spectrum lines situated in different spectral regions. These regions are shown on Fig. \ref{fig:ccd_rgb} and are provisionally denoted as blue, green and red. Figure \ref{fig:fwhm_cols} shows the FWHMs distribution along columns (dispersion). No significant trends either along columns or rows can be seen, what means that the focal plane of the lens is well-aligned with the plane of the CCD chip.   

\begin{figure}[!htb]
  \begin{center}
    \includegraphics[width=0.75\textwidth, trim = 0mm 0mm 0mm 0mm, clip=true]{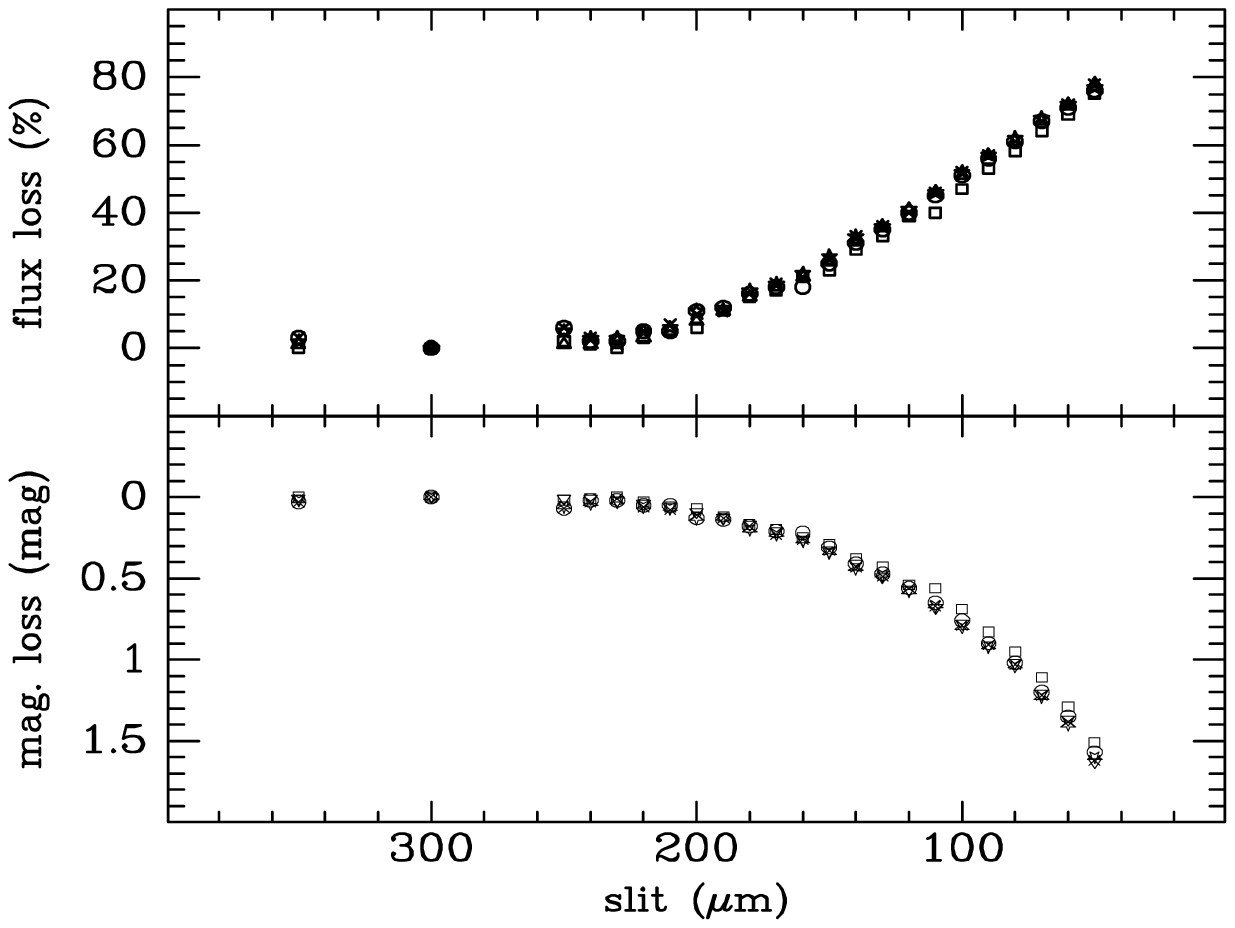}
    \caption[]{Actual signal loss when narrowing the slit in terms of flux (upper panel) and stellar magnitudes (down). Flux loss is measured in percentages according the signal of full opened slit. The uncertainties of the flux and of the corresponding magnitude loss  are presented in Tab. \ref{tab:losses} in columns (3) and (5), respectively.}
    \label{fig:flux_mag_loss}
  \end{center}
\end{figure}

To evaluate the signal loss at smaller slit width twenty three thorium-argon lamp echellograms were taken, gradually changing the width of the slit from $350 \mu m$ to $50 \mu m$, and keeping the exposure constant. For every echellogram (corresponding to a certain slit width as well) a single (unblended) unsaturated thorium-argon emission line was Gaussian approximated within five randomly chosen orders. This way twenty three flux values corresponding to different slit widths were derived. The change of these flux values along echelograms was accepted as a measure of flux loss related to the decrease of the slit width. 
For every emission line Gaussian flux parameter values were normalized to maximum flux value in that sample ($flux_{max}$). The normalized flux values in percentage corresponding to the five lines in every echellogram were averaged to give a mean flux value at every slit width. The mean flux standard deviation was calculated also and its value accepted as an error of the flux loss measurement. To every flux loss value we attached a magnitude, following $-2.5log(flux/flux_{max})$. These data are listed in Tab. \ref{tab:losses}. Column (1) presents the slit width, columns (2) and (3) contain the flux loss and its uncertainty in percentage, columns(4) and (5) present the same in terms of stellar magnitudes as defined above, columns (6) and (7) contain the FWHM of the Gaussian fit  and its uncertainties.
\begin{table}[h!tb]
\begin{center}
  \caption{Flux and magnitude losses caused by decrease of the slit width}
  \begin{tabular}{cccccccc}
slit &  &	flux    &   stddev      &  magnitude    &  stddev & fwhm & fwhm err  \\
um & & \% & \% & mag & mag & px & px  \\
      \hline\\
  50 &  &  76.8   &    1.3      &        1.59   &    0.05  &   1.69  &     0.08      \\
  60 &  &  71.2   &    1.3      &        1.36   &    0.04  &   1.82  &     0.08      \\
  70 &  &  66.6   &    1.5      &        1.19   &    0.05  &   1.94  &     0.05      \\
  80 &  &  60.6  &     1.5     &         1.01  &     0.04  &   2.08  &     0.04     \\
  90 &  &  56.0   &    1.7      &        0.89   &    0.04  &   2.13  &     0.07      \\
 100 &  &  50.6   &    2.1      &        0.76   &    0.04  &   2.28  &     0.05      \\
 110 &  &  44.6   &    2.6      &        0.65   &    0.05  &   2.39  &     0.07      \\
 120 &  &  40.2   &    0.8      &        0.56   &    0.01  &   2.53  &     0.07      \\
 130 &  &  35.2  &     1.3     &         0.47  &     0.02  &   2.70  &     0.04     \\
 140 &  &  31.6  &     1.7     &         0.41  &     0.02  &   2.83  &     0.05     \\
 150 &  &  25.4  &     1.5     &         0.32  &     0.02  &   2.95  &     0.06     \\
 160 &  &  20.6  &     1.5     &         0.25  &     0.02  &   3.08  &     0.05     \\
 170 &  &  18.2  &     0.8     &         0.22  &     0.01  &   3.21  &     0.05     \\
 180 &  &  15.8  &     0.8     &         0.18  &     0.01  &   3.29  &     0.04     \\
 190 &  &  11.2  &     0.4     &         0.13  &     0.01  &   3.43  &     0.04     \\
 200 &  &   9.2  &     2.2     &         0.11  &     0.03  &   3.54  &     0.05     \\
 210 &  &   5.6  &     0.9     &         0.06  &     0.01  &   3.63  &     0.06     \\
 220 &  &   4.2  &     1.1     &         0.05  &     0.01  &   3.67  &     0.06     \\
 230 &  &   1.6  &     1.1     &         0.02  &     0.01  &   3.77  &     0.04     \\
 240 &  &   2.0  &     1.0     &         0.02  &     0.01  &   3.79  &     0.07     \\
 250 &  &   4.0  &     2.3     &         0.05  &     0.03  &   3.81  &     0.07     \\
 300 &  &   0.0  &     0.0     &         0.00  &     0.00  &   3.87  &     0.08     \\
 350 &  &   1.8  &     1.3     &         0.02  &     0.01  &   3.84  &     0.10     \\
\end{tabular}
\label{tab:losses}
\end{center}
\end{table} 
The upper panel in Fig. \ref{fig:flux_mag_loss} shows the flux loss (in percentages) vs slit width, and the lower panel shows the flux loss in terms of magnitude.

\begin{figure}[!htb]
  \begin{center}
    \includegraphics[width=0.70\textwidth, trim = 0mm 0mm 0mm 0mm, clip=true]{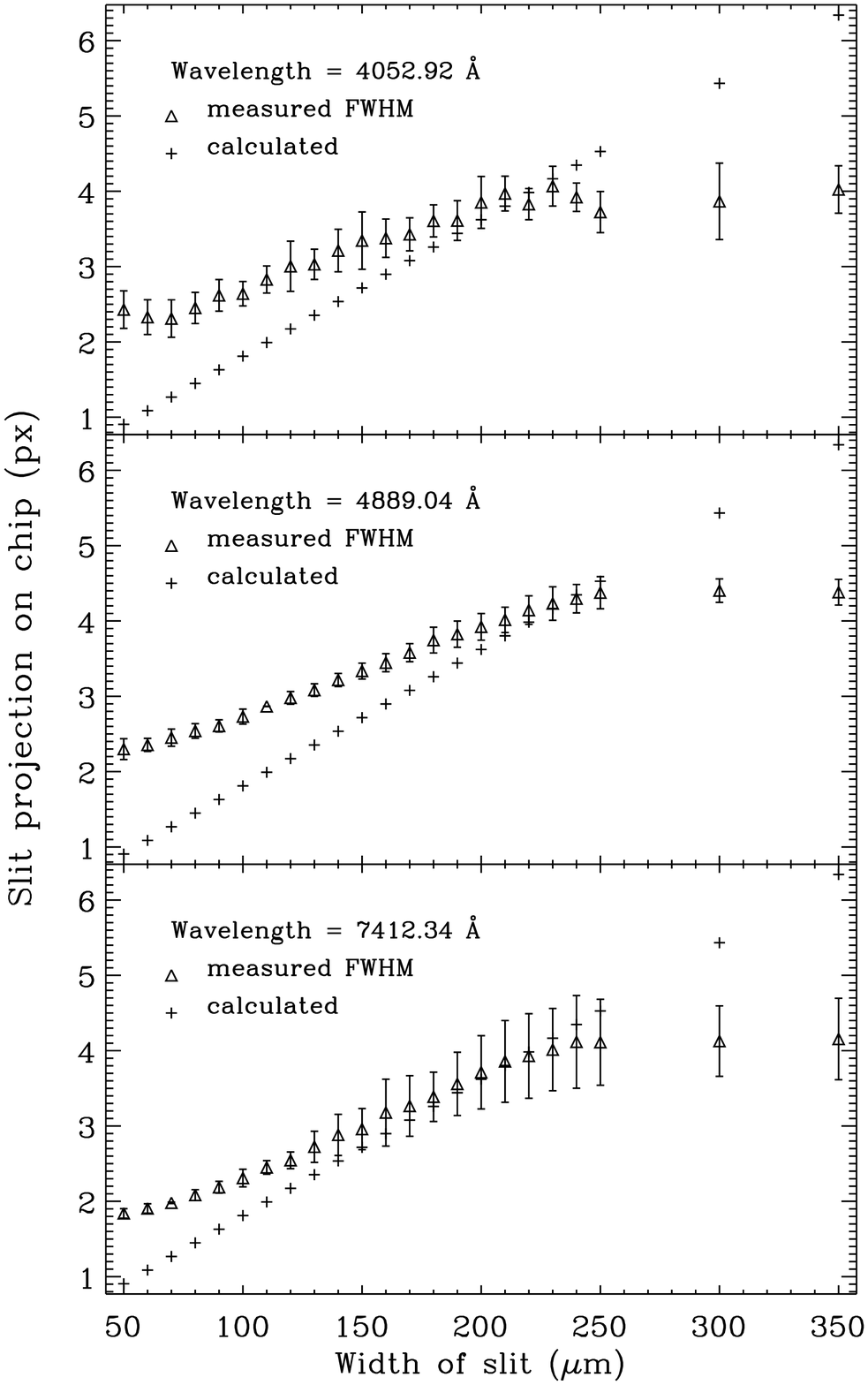}    
    \caption[]{Dependence of the FWHM on the slit width for three selected ThAr lines (marked with circles in fig. \ref{fig:ccd_rgb}).}  
    \label{fig:slit_project}
  \end{center}
\end{figure}

Another illustration of a possible change of the FWHM of the spectral lines with variations of the slit width is presented on Fig. \ref{fig:slit_project}. Here, we use only the three lines marked on Fig. \ref{fig:ccd_rgb}. A  subimage ($\pm$7 x $\pm$5 px along the horizontal and vertical direction, respectively) is extracted around the center of every line, and transformed to 1D, by averaging along the vertical direction. The 1D-line was fitted with a Gauss function, and the FWHM was derived from the parameters of the fit. The comparison between the measured FWHM and the calculated projection of the slit width shows that no increase of the FWHM is experienced for slit widths greater than about 220 $\mu$m. This can be considered as a measure of the diameter of the beam image in the focal plane of the collimator (the plane of the slit). In the region of very small slit widths the FWHM, as expected, approximates gradually the limit of 2 px (Nyquist theorem).  

\begin{figure}[h!tb]
  \begin{center}
    \includegraphics[width=0.7\textwidth, trim = 0mm 0mm 0mm 0mm, clip=true]{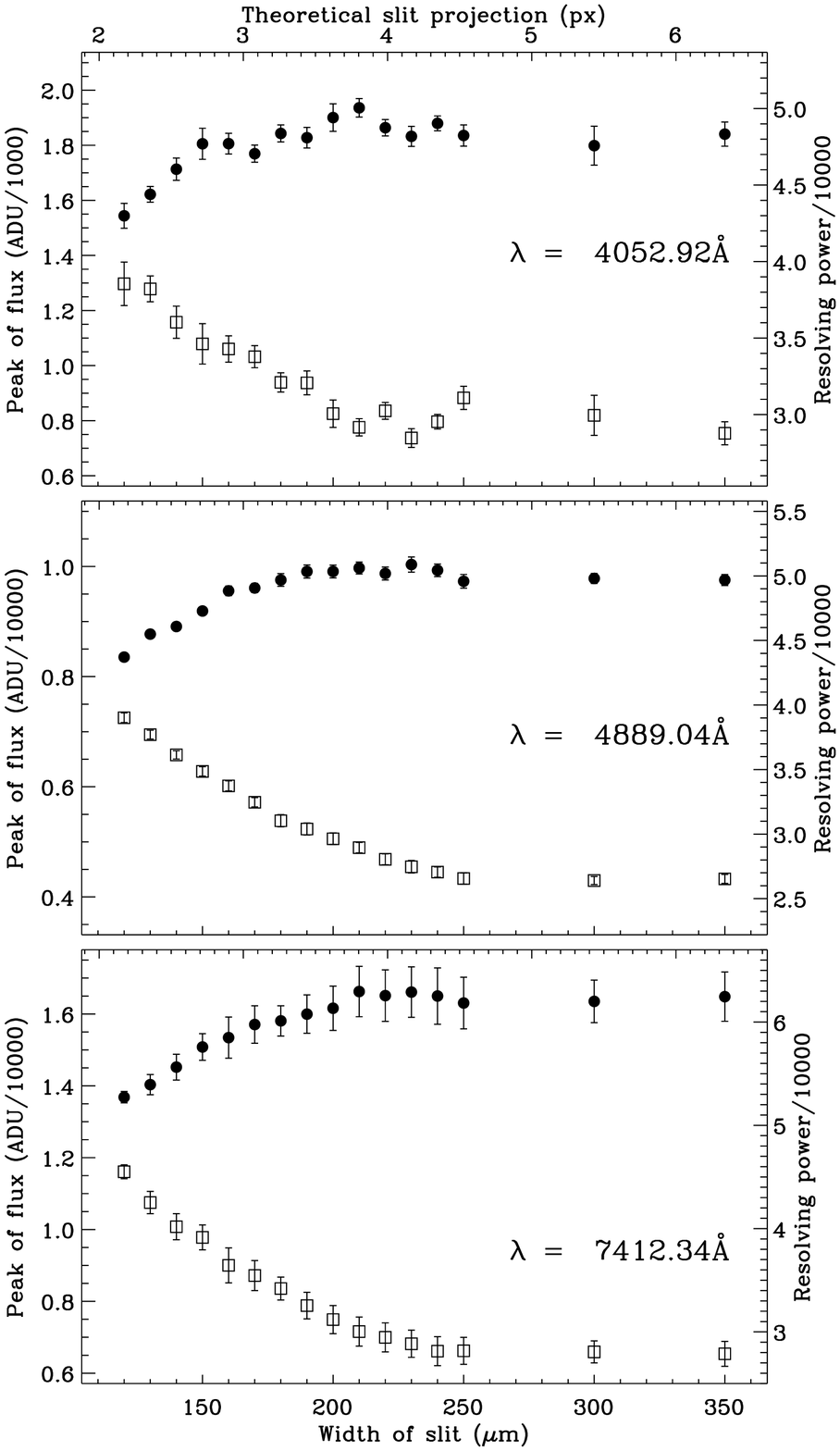} 
    \caption{
Flux and resolving power versus slit width. The lower abscissa shows the slit width and the upper one the size of its projection on the CCD chip.    
Filled circles mark the maxima of the fitted profiles, and their values are on the left ordinate. The resolving power is marked with empty squares, and their values can be read on the right ordinate.}   
    \label{fig:resolve_power}
  \end{center}
\end{figure}

The measurements of the three lines marked on Fig. \ref{fig:ccd_rgb} were further used to assess together both, the increase in resolving power and the decrease in flux when narrowing  the slit width. These dependencies are presented on Fig. \ref{fig:resolve_power}. As a measure of the decreased flux we used the maxima of the Gaussian fit, described in the previous paragraph. The scale for the resolving power is on the right ordinate. For slit widths greater than about 200 $\mu$m no substantial changes of the flux and resolution are observed, because this is equivalent to taking spectra without slit. For slit widths lower than about 110 $\mu$m (corresponding to $\approx$2 px projection on the CCD) the flux losses dominate over the increase in resolution, thus it is not recommended to close the slit below this limit.   

\subsection*{3.3. Signal-to-noise ratio}

\begin{figure}[!htb]
  \begin{center}
    \includegraphics[width=0.7\textwidth, trim = 0mm 0mm 0mm 0mm, clip=true]{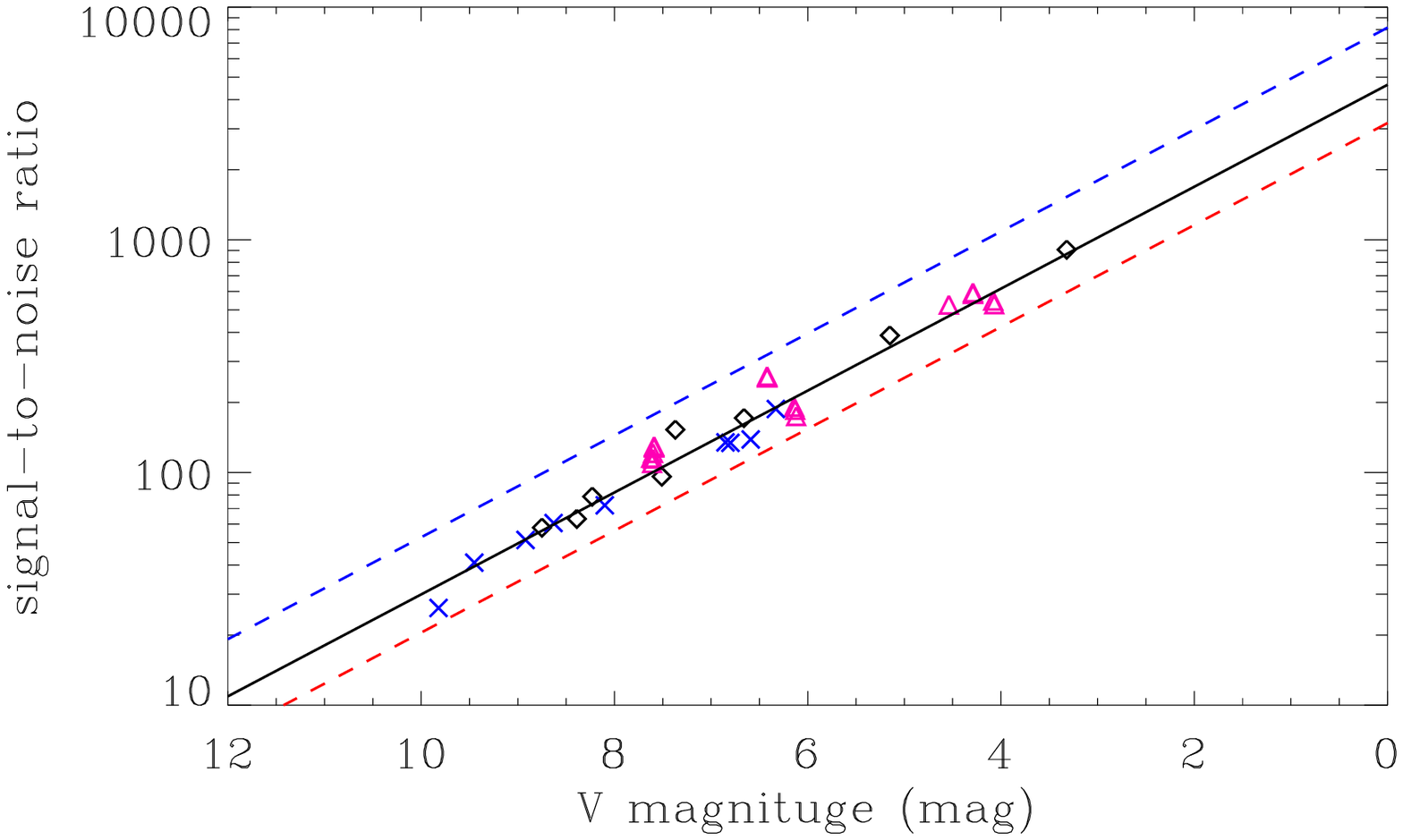}
    \caption[]{Signal-to-noise ratio vs. V magnitude.}
    \label{fig: snr-mag}
  \end{center}
\end{figure}

A good measure of the efficiency of the spectrograph is the Signal-to-noise ratio (SNR) obtained at different illumination levels. Figure \ref{fig: snr-mag} shows the derived SNR per pixel as a function of the stellar magnitude. This relation is a compilation of three clear observing nights with more than 8 observed objects per night and seeing 1-2\arcsec. The seeing is evaluated from  stellar image projection on the fiber entrance placed in the telescope focal pane. The SNR measurements of the nights with seeing 1 - 1.5\arcsec were transformed to the results from the night with 2\arcsec seeing using least square method. Stars of different magnitudes were observed with different exposures, but for the final set of data they were all transformed to exposure of 1800 s. The straight line presents linear regression taken from the SNR measurements of the three nights. Spectra taken at airmass $>$ 2 were not used. Stellar magnitudes of the objects were reduced to unity airmass using extinction coefficient of 0.2, a value commonly accepted at NAO Rozhen. The dashed lines in Fig. \ref{fig: snr-mag}, above and below the empirically derived dependence, show the expected signal-to-noise ratio in case of seeing 1\arcsec and 3\arcsec, respectively. These estimations were made with a 2D Gaussian function, calculated for different seeing sizes, and measuring the portion of the flux proportional to the cross-section of the entrance tip of the object fiber. From the data summarized in Fig. \ref{fig: snr-mag}, for SNR=100 and exposure 1800 s, the limiting magnitudes are 8.8, 7.5, and 6.8, for seeing 1\arcsec, 2\arcsec, and 3\arcsec, respectively.  

\section*{3.4.Radial velocity accuracy measurement}

\begin{figure}[!htb]
  \begin{center}
    \includegraphics[width=0.75\textwidth, trim = 0mm 0mm 0mm 0mm, clip=true]
{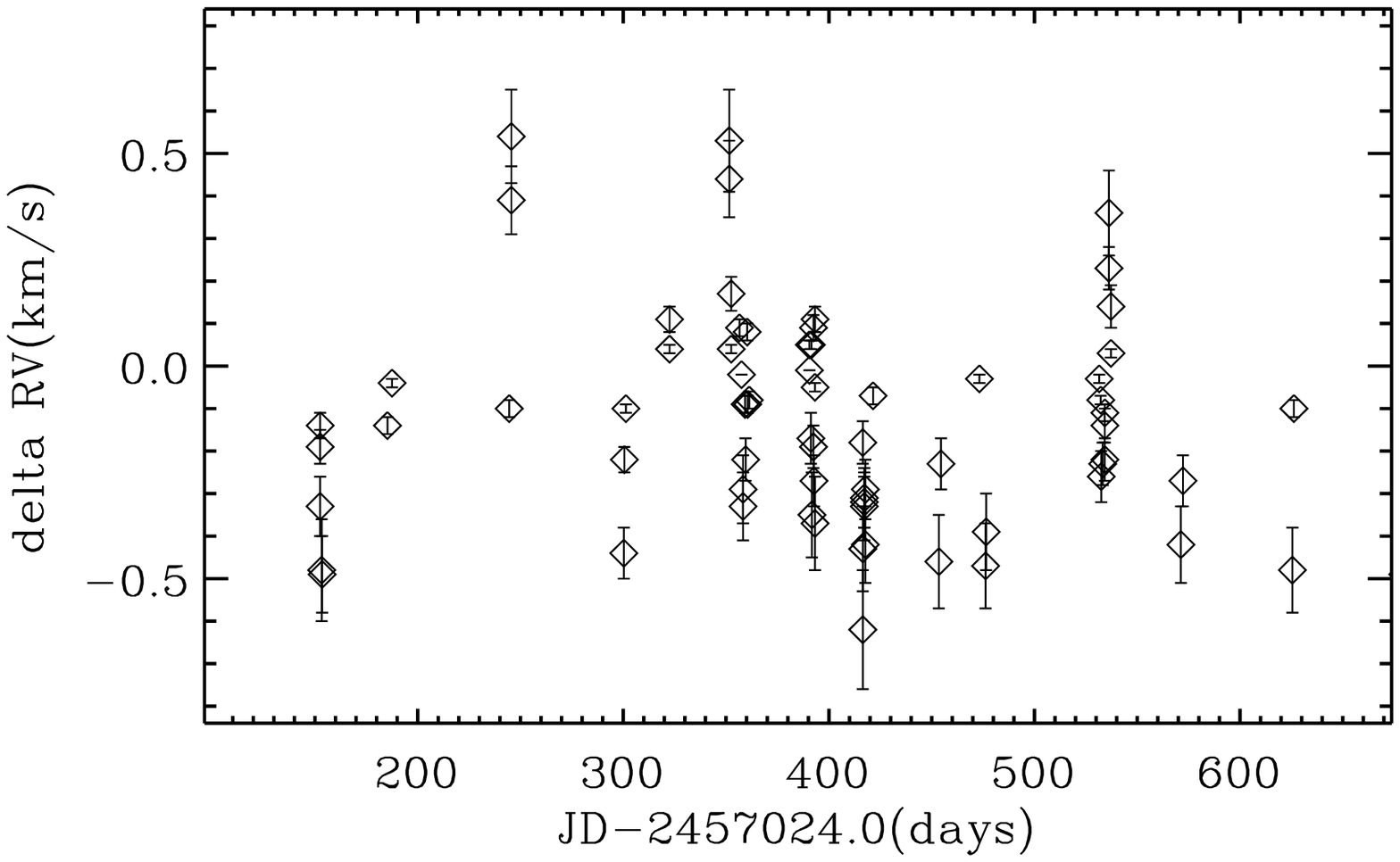}
    \caption[]{Radial velocity measurments of RV standard stars in the period 2 Jun 2015 - 18 Sep 2016.}
    \label{fig:rv-monitor}
  \end{center}
\end{figure}

In order to determine the zero point of the radial velocities (RV) derived from ESpeRo spectra observations of appropriate standard stars were obtained every observing night. Stars were selected mainly from the GAIA catalog of RV standard stars (Soubiran et al., 2013). Sixty nine spectra of twenty three standard stars were obtained in the period June 2015 - September 2016. RVs were obtained by crosscorrelating the spectrum of the studied object with a template spectrum (template). Templates were generated using SPECTRUM, a stellar spectral synthesis program written by Gray and Corbally (1994), and POLLUX, a stellar spectra database offering access to theoretical data (Palacios et al., 2010). Every RV is derived as a mean value of measurements obtained from many orders. Orders with a low signal-to-noise ratio as well as strong telluric absorption features were omitted. Figure \ref{fig:rv-monitor} presents the difference $catalog - measured$ radial velocity value. The annual monitoring of this difference shows slight systematic deviation of measured velocities with a mean value of -0.18 km s$^{\rm -1}$ with a standard deviation of the mean 0.07 km s$^{\rm -1}$. More thorough analysis of the radial velocity accuracy  derived from ESpeRo and its dependence on spectrograph working conditions will be discussed in a forthcoming paper.

\subsection*{3.5 Sample spectra}
\begin{figure}[!htb]
  \begin{center}
    \includegraphics[width=0.7\textwidth, trim = 0mm 0mm 0mm 0mm, clip=true]{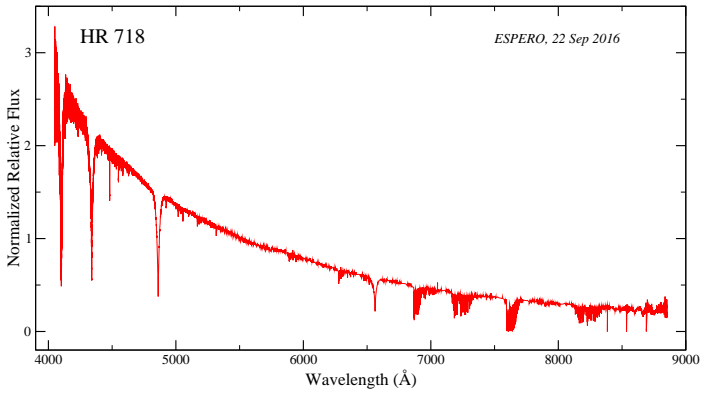}
    \caption[]{Spectrum of HR 718 calibrated to relative fluxes,  normalized at 5500 {\AA} .}
    \label{fig:hr718_flux}
  \end{center}
\end{figure}

The coverage of the full spectral range of the spectrograph is best illustrated after merging all orders into a single 1-D spectrum. Here again, as described in section \ref{subsec:identi_ord}, the good flat field for every order is very important, to remove the blaze function of the echelle grating. Some residuals after dividing the orders by their corresponding flat fields can be removed by using spectra of standard stars. Spectrophotometric standard-star observations are also needed to derive the response function of the spectrograph. One example of a 2-D echellogram transformed into an 1-D spectrum, and calibrated in relative fluxes is given in Fig. \ref{fig:hr718_flux}.  

\begin{figure}[!htb]
  \begin{center}
    \includegraphics[width=0.7\textwidth, trim = 0mm 0mm 0mm 0mm, clip=true]{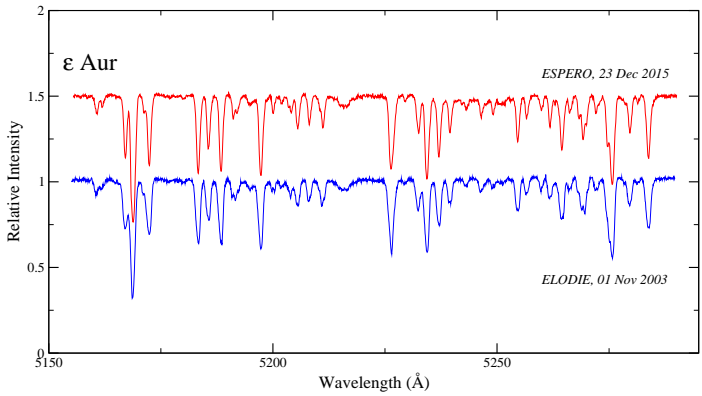}
    \caption[]{Comparison of extracted portion of spectra of the star $\varepsilon$ Auriga, obtained with ESpeRo and Elodie.}
    \label{fig:epsaur_comp}
  \end{center}
\end{figure}

Given the described characteristics of the spectrograph it is interesting to compare spectra of ESpeRo with spectra of the same objects, obtained with echelle spectrograph used on a 2-meter class telescope. In Fig. \ref{fig:epsaur_comp} part of the spectrum of $\varepsilon$~Aur from the ELODIE database (Moultaka et al. 2004) is compared with one obtained with ESpeRO.
The nominal resolution of Elodie is about 42000, and the spectrum obtained with ESpeRo is exposed without slit, i.e. the R is about 30000. Nevertheless no substantial differences are seen between both spectra.  

\section*{4. First scientific results}
The first published result, based on data obtained with ESpeRo is presented in Bonev et al. (2014). In this Telegram the authors report on the most prominent features measured in a spectrum of SN2014J on January 23.75 UT, 2014.  Stoyanov and Zamanov (2016) present optical spectroscopy of the high-mass X-ray Binary A0535+26 (HDE 245770, V725 Tau) after the periastron passage. 
Zamanov et al. (2016) describe spectral observations of four Be/X-ray and
$\gamma$-ray binaries. Tomov et al. (2016) studied the symbiotic star AG~Peg during its 2015 outburst.

\section*{Summary}
ESpeRo is now operational. Here we presented its design, performance and working status. Several improvements can enhance its efficiency in future. So, e.g. a focal reducer could be used in the RC-focus for better matching between the typical seeing at Rozhen and the parameters of the fiber injection unit.
Alternative methods for obtaining flat fields better matching the science frames (one option is to design  a system for obtaining dome flats). With a specially designed filter the illumination of the tungsten lamp (and LED) could be modified to have a rather flat spectral distribution. New algorithms for more efficient autoguiding are under development. Replacement of the cross-disperser with a new one aiming to improve light transmission in the blue spectral range. A new software dedicated to control spectrograph functions and to complete FITS headers with a relevant astronomical information during the observing night is still needed.

{\em Acknowledgements:} We are thankful to Remi Cabanac and his team at Pic du Midi observatory and Tarbes for the extensive and valuable discussions concerning the design of the Narval spectrograph, including its budget and prices of components. In Toulouse, Laurent Pares provided us very detailed information about the optical design of echelle spectrographs in general, and about ESPaDOnS in particular. The Plant For Optics\footnote{http://www.pfo-bg.com/} in Panagyurishte played a very important role in the process of assembling and alignment of the spectrograph. 
This research made use of the POLLUX database\footnote{ http://pollux.graal.univ-montp2.fr} operated at LUPM  (Université Montpellier - CNRS, France with the support of the PNPS and INSU.).  The preliminary processing and the data analysis was performed mainly with IRAF routines \footnote{IRAF is distributed by the National Optical Astronomy Observatories, which are operated by the Association of Universities for Research in Astronomy, Inc., under cooperative agreement with the National Science Foundation}. 
High resolution stellar spectra were downloaded from the ELODIE archive\footnote{http://atlas.obs-hp.fr/elodie/} in order to be compared with spectra obtained with ESpeRo.
The development of the design of the spectrograph and its manufacturing were financially supported by contracts DO 02-85 and DO 1-34 with the Ministry of science and education. The authors would like to thank the the Bulgarian Academy of Sciences for the support for contracts DFNP-103/11.05.2016, DFNP-106/11.05.2016, DFNP-107/11.05.2016, DFNP-122/11.05.2016 "Support for young scientists".


\end{document}